\documentclass[12pt]{article}

\input{myCommands.sty}
\usepackage[colorlinks,citecolor=blue,urlcolor=blue,filecolor=blue,backref=page]{hyperref}

\newcommand{\blind}{0}

\newcommand{\beginsupplement}{%
        \setcounter{section}{0}
        \renewcommand{\thesection}{S\arabic{section}}
        \setcounter{table}{0}
        \renewcommand{\thetable}{S\arabic{table}}%
        \setcounter{figure}{0}
        \renewcommand{\thefigure}{S\arabic{figure}}%
        \setcounter{page}{1}
        \renewcommand{\thepage}{S\arabic{page}}
        \setcounter{equation}{0}
        \renewcommand{\theequation}{S\arabic{equation}}
     }

\begin{document}

\def\spacingset#1{\renewcommand{\baselinestretch}%
{#1}\small\normalsize} \spacingset{1}


\if0\blind
{
  \title{\bf Bayesian Nonparametric Density Autoregression with Lag Selection}
  \author{Matthew Heiner and 
Athanasios Kottas\thanks{
{M. Heiner is Assistant Professor, Department of Statistics,
Brigham Young University, and A. Kottas is Professor, Department of Statistics, University of California, Santa Cruz. This work is part of the Ph.D. dissertation of the first author, completed at the University of California, Santa Cruz. The authors wish to thank Stephan Munch for several useful discussions.} This research was supported in part by the National Science Foundation under award SES 1631963.}\hspace{.2cm}} 
  \maketitle
} \fi

\if1\blind
{
  \bigskip
  \bigskip
  \bigskip
  \begin{center}
    {\LARGE\bf Bayesian Nonparametric Density Autoregression with Lag Selection}
\end{center}
  \medskip
} \fi

\bigskip
\begin{abstract} 
We develop 
a Bayesian nonparametric autoregressive model applied to flexibly estimate general transition densities exhibiting nonlinear lag dependence. Our approach is related to Bayesian 
density regression using Dirichlet process mixtures, with the Markovian likelihood defined 
through the conditional distribution obtained from the mixture. This results in a Bayesian nonparametric extension of a mixtures-of-experts model formulation. We address computational challenges to posterior sampling that arise from the 
Markovian structure in the likelihood. The base model is illustrated with synthetic data from a classical model for population dynamics, as well as a series of waiting times between eruptions of Old Faithful Geyser. We study inferences available through the base model before extending the methodology to include automatic relevance detection among a pre-specified set of lags. Inference for global and local lag selection is explored with additional simulation studies, and the methods are illustrated through analysis of an annual time series of pink salmon abundance in a stream in Alaska. We further explore and compare transition density estimation performance for alternative configurations of the proposed model. 
\end{abstract}

\noindent%
\vfill

\newpage
\spacingset{1.5} 

\section{Introduction}
\label{sec:intro}
 
This article is concerned with flexible transition density estimation for non-stationary, nonlinear time series. Let $\{y_t : t = 1, \ldots, T \}$ denote a univariate series governed by a time-homogeneous transition density $p(y_t \mid y_{t-1}, \ldots, y_1)$. While nonlinearity has been used to describe various qualitative characteristics of time series, we specifically refer to nonlinear dynamics, or the function mapping past observations to the present. Many existing methods for nonlinear regression have been applied to autoregressive modeling within and out of the statistical literature. Density regression has received far less attention, especially in application to transition density estimation, a crucial component of probabilistic forecasting and decision modeling. We seek to build on recent advances in transition density estimation by 
exploring what can be succinctly described as \color{black} an extension to \color{black} Bayesian nonparametric mixtures of autoregressive models. To accommodate nonlinear dependence, mixture weights are functions of lagged observations. Thus, our method is also accurately described as a locally linear autoregressive model.



Perhaps the most popular mixture modeling application to time series is the class of hidden Markov models (HMMs), which are capable of capturing nonlinear dynamics (\citealp{fruhwirth2006}, and references therein). Markovian dependence in a latent process, however, complicates inferences for transition densities and related functionals, especially when considering multiple lags. \color{black} The likewise popular classes of threshold autoregressive models \citep{tong1990tar}, and mixtures-of-experts (MoE) models \citep{jordan1994hme, peng1996bayesmoe, carvalho2005, carvalho2006modeling} alternatively build dependence into mixture weights through lagged observations directly. We take the MoE approach, replacing parameterized link functions of lagged observations with normalized kernels for local weighting \citep{glasbey2001gmar, kalliovirta2015gmar}.

\color{black}




In contrast with \color{black} most HMM and MoE methods\color{black}, our \color{black} models are based on \color{black} countable mixtures, bypassing the need to \color{black} fix the number of mixture \color{black} components. 
\color{black} Bayesian nonparametric (BNP) approaches have expanded the \color{black} 
hidden Markov \citep{beal2002, taddy2009markov, yau2011bnpHMM} and dynamic linear \citep{rodriguez2008dynamicDensity, caron2007dpdlm, fox2011dpDLM} model frameworks.
Dirichlet process mixtures (DPM; \citealp{ferguson1973,antoniak1974mixtures}) of linear autoregressive (AR) models \citep{lau2008dpmar, dilucca2013}, 
which are closer to our formulation, 
can be viewed as nonparametric extensions of the mixture autoregressive model of \citet{wong2000mixture}. 
\color{black}
\color{black} 
DPM of AR models typically use static weights, restricting transition mean functionals to be linear. 
\color{black}
\citet{muller1997} 
use normalized weights that employ a finite MoE framework to accommodate nonlinearity.
Posterior consistency for 
BNP transition density estimation has been explored by \citet{tang2007consistent}, \citet{tang2007transdens}, and \citet{chae2019}.


\color{black}
Many of the above methods assume first-order time dependence. While convenient and occasionally justified, this assumption may over-simplify or misspecify the dynamics. 
Higher-order models can also enable 
phase-space reconstruction via time-delay embedding. Although applied to deterministic systems, a theorem by \citet{takens1981} justifies reconstructing multidimensional dynamical systems, up to topological equivalence, using only lags of a univariate time series. Markovian stochastic models can approximate this method for applications that exhibit noise \citep[Ch.\ 10, 12]{kantz2004book}. The practical utility of this result is evident in fields like ecology, where full observation of all relevant variables is practically impossible. 
\color{black}

\color{black}
Motivated by these considerations, we propose to model
\color{black}
the transition density for observation $y_t$, conditional on $L$ lags $\bm{y}_{t-1:L} \equiv (y_{t-1}, \ldots, y_{t-L})$, as $ \sum_{h=1}^\infty q_h(\bm{y}_{t-1:L}) \, p_h( y_t \mid \bm{y}_{t-1:L} )$, with component-specific normalized weight functions $q_h( \bm{y}_{t-1:L} )$ and kernel densities 
$p_h( y_t \mid \bm{y}_{t-1:L} )$. 
\color{black} The model form resembles that of \color{black} 
\citet{antoniano2016}, who build on \citet{martinez2011time}, constructing a transition density from a mixture model on the stationary joint density of the current observation and a single lag. 
Their likelihood is based on the conditional transition density, which is a nonparametric mixture of kernels with linear autoregressive means and lag-dependent, normalized weights. \citet{kalli2018npbvar}
extend this framework to a stationary multivariate autoregressive model of multiple lags, although 
the model is implemented with a single lag. \citet{deyoreo2017} use a similar model construction, 
\color{black}
achieving superior flexibility by relaxing the stationarity assumption. \color{black} 
The model proposed in this article extends that of \citet{deyoreo2017} to accommodate multiple lags and, crucially, shrink dependence to a minimally sufficient set of lags. The added modeling and computational complexity associated with high-order dependence demands that lag selection play a vital role in this work, as it affords parsimony and significantly reduces the estimation burden.

The primary contributions of this article are 1) extension of a powerful class of non-stationary, nonlinear density autoregression models to accommodate dependence on multiple lags; 2) development of a framework for model-based selection and exploration of lag dependence; 3) investigation into the proposed model's fitness for different analysis scenarios; and 4) demonstration of the need for lag selection in high-order density autoregression.

\textcolor{black}{The rest of the article is organized as follows.}
In Section \ref{sec:model}, we propose a BNP time-series model for density autoregression and present details for implementation and inference. In Section \ref{sec:illustrations}, we illustrate the model fit to synthetic and real data. In Section \ref{sec:lagselection}, we extend the model to incorporate inferences about relevant lags and demonstrate its use on data. Section \ref{sec:forec_comparison} compares transition density estimation performance 
\textcolor{black}{under different model configurations, using}
simulated nonlinear time series featuring skewness, heteroscedasticity, and different lag dependence structures.
\textcolor{black}{Finally, Section \ref{sec:discussion} concludes with discussion.
The Supplementary Material contains details on: model modifications for stationary time series; prior specification; \color{black} computing time and sensitivity analysis; \color{black} the Markov chain Monte Carlo (MCMC) algorithms for the base model and its extension that incorporates lag selection; and an additional simulation example.}

\section{\textcolor{black}{The modeling approach}}
\label{sec:model}

Our objective is to develop a general-purpose and fully nonparametric, time-homogeneous Markovian model that is sufficiently flexible to: 1) estimate possible non-Gaussian transition densities, dependent on lagged values, 2) capture nonlinear dynamics, and 3) select relevant lags among a pre-specified set, up to a maximal order $L$. The first two objectives are accomplished through a nonparametric mixture of Gaussian densities, wherein both the mixture weights and kernel means depend on lagged observations. 
The general model formulation for the transition density can be written as
\begin{align}
  \label{eq:generalmodel}
  f(y_t \mid \bm{y}_{t-1:L}) = \sum_{h=1}^\infty \underbrace{ q_h(\bm{y}_{t-1:L}) }_{\text{local weights}} \underbrace{\Ndist( y_t \mid \mu_h(\bm{y}_{t-1:L}), \sigma_h^2 )}_{\text{mixture kernels}} \, ,
\end{align}
where $\Ndist(y \mid \mu, \sigma^2)$ denotes a Gaussian density with mean $\mu$ and variance $\sigma^2$ evaluated at $y$, and with weight function $q_h(\bm{y}_{t-1:L}) \ge 0 $ for all $h \in \mathbb{N}$ such that $\sum_{h=1}^\infty q_h(\bm{y}_{t-1:L}) = 1$ for all $\bm{y}_{t-1:L} \in \mathbb{R}^L$. \color{black} We utilize \color{black} kernel mean functions $\mu_h(\bm{y}_{t-1:L})$ \color{black} that \color{black} are linear in the lags, yielding a local linear model formulation. The objective of order and lag selection is accomplished through a stochastic-search prior structure.

Time homogeneity 
is a consequence of time invariance in the parameters governing the mixture weights and kernels. We note that this seemingly restrictive assumption is at least partially offset by the model's flexibility with respect to lagged observations. Apparently time-dependent structural changes can sometimes be attributed to heterogeneity of response across the state space. In such cases, a latent first-order Markov process governing the mixture weights may be less effective than our approach of using the lagged values directly. Nevertheless, dynamic drift or regime-switching in model structure may be more appropriate in some scenarios, for which we urge thoughtful exploration before selecting a model. 

\textcolor{black}{We proceed with model specification in Section \ref{sec:modelSpec}, built using a covariance matrix parameterization that is useful for interpretation and implementation.}
Section \ref{sec:prior_settings} discusses the roles of model parameters and gives recommended prior settings. Section \ref{sec:computation} briefly outlines the MCMC algorithm used for posterior inferences and addresses implementation. Finally, Section \ref{sec:transdens_estimation} discusses model inferences, including transition density estimation. 

\subsection{Model specification}
\label{sec:modelSpec}

One avenue to arrive at the conditional density form in (\ref{eq:generalmodel}) begins with a prior for joint density estimation. For clarity in notation, we use $y \in \mathbb{R}$ to represent the current observation and $\bm{x} \equiv \bm{y}_{t-1:L} \in \mathbb{R}^L $ to denote the lags. We begin as in \citet{muller1996curve}, who in the regression setting consider $y$ and $\bm{x}$ to arise jointly from a Gaussian DPM. 
This implies the stick-breaking representation \citep{sethuraman1994} for joint density,
\begin{align}
  \label{eq:jointmodel_stickbreak}
  f_{YX}(y,\bm{x} \mid G) = \sum_{h=1}^\infty \omega_h \Ndist \left( (y,\bm{x}) 
		\mid \bm{\mu}_h, \bm{\Sigma}_h \right) \, ,
\end{align}
where 
\textcolor{black}{the $(\bm{\mu}_h,\bm{\Sigma}_h)$ arise i.i.d. from the Dirichlet process (DP) centering distribution $G_{0}$, and the mixture weights, $\omega_h$,} are constructed as
\begin{align}
	\label{eq:stickbreaking}
	\omega_1 = v_1, \ \omega_h = v_h \prod_{j=1}^{h-1} (1 - v_j), \ \text{for}\ h > 2, \ \text{and} \ v_h \simiid \Betadist(1, \alpha) \, .
\end{align}
Conditioning on $\bm{x}$, we obtain
\textcolor{black}{the transition density model:}
\begin{align}
  \label{eq:condmodel}
  f_{Y \mid X}(y \mid \bm{x},G) 
  = \frac{ \sum_{h=1}^\infty \omega_h\Ndist_{(h)}(\bm{x})  \Ndist_{(h)} ( y \mid \bm{x} ) }
  { \sum_{j=1}^\infty \omega_j \Ndist_{(j)}(\bm{x}) } 
  = \sum_{h=1}^\infty q_h(\bm{x}) \Ndist( y_t \mid \mu_h(\bm{x}), \sigma_h^2 ) \, ,
\end{align}
with $q_h(\bm{x}) = \omega_h \Ndist_{(h)}(\bm{x}) / \sum_{j=1}^\infty \omega_j \Ndist_{(j)}(\bm{x})$, where $\Ndist_{(h)}(\cdot)$ refers to a Gaussian density with parameters corresponding to mixture component $h$, and $\Ndist_{(h)}(y \mid \bm{x})$ is the univariate conditional Gaussian density derived from $\Ndist_{(h)}(y, \bm{x})$. The joint densities in each mixture component of the numerator of (\ref{eq:condmodel}) have been factored into their respective marginal $L$-dimensional Gaussian density for $\bm{x}$ (with mean $\bm{\mu}^x$ and covariance $\bm{\Sigma}^x$) and univariate conditional Gaussian density for $y$ (with \color{black} linear \color{black} mean $\mu(\bm{x}) \equiv \mu^y + \bm{\Sigma}^{yx} (\bm{\Sigma}^{x})^{-1}(\bm{x} - \bm{\mu}^x)$ and variance $\sigma^2 \equiv (\sigma^y)^2 - \bm{\Sigma}^{yx} (\bm{\Sigma}^{x})^{-1} \bm{\Sigma}^{xy}$). The second line of (\ref{eq:condmodel}) reveals the local linear model structure with lag-dependent weights. 

This procedure yields 
a conditional density that satisfies the requirements of the proposed model (\ref{eq:generalmodel}). Specifically, 
\textcolor{black}{since $\sum_{h=1}^\infty \omega_h = 1$ almost surely,}
so long as there exists some positive constant $c_N < +\infty$ such that $0 < \Ndist_{(h)}(\bm{x}) < c_N$ for all $h \in \mathbb{N}$ and all $\bm{x} \in \mathbb{R}^L$ (which is satisfied if there exists another constant $c_\Sigma > 0$ such that $\det( \bm{\Sigma}_{h}^x ) > c_\Sigma $ for all $h \in \mathbb{N}$), the denominator in $q_h(\bm{x})$ will be positive and finite for all $\bm{x} \in \mathbb{R}^L$.

Although $\bm{x}$ (representing $\bm{y}_{t-1:L}$) can legitimately be considered random in the time-series context, the Markovian likelihood 
requires that the conditional density (\ref{eq:condmodel}) form the basis of the model. 
\color{black}
Besides creating redundancy in the likelihood, modeling separate joint distributions for consecutive length-$(L+1)$ coordinate vectors would not generally be coherent. To see this, consider $y_t$, which appears in both $(y_{t+1}, \bm{y}_t)$ and $(y_t,\bm{y}_{t-1:L})$. Modeling each vector with a joint mixture as in \eqref{eq:jointmodel_stickbreak} would result in two distinct marginal distributions for $y_t$ without additional assumptions, like strong stationarity. 
We forego stationarity in favor of flexibility.
Consequently, we interpret the $ \{ \Ndist_{(h)}(\bm{x}) \} $ densities in $\{ q_h(\bm{x}) \}$ {\em exclusively} as functions that localize the mixture weights, and not as joint densities of lagged observations. Indeed, localizing the weights is their {\em only} role in a conditional likelihood based on \eqref{eq:condmodel}. \color{black}
Supplement \ref{sec:appendix_stationary} includes discussion of possible mixture model formulations for the stationary case.
\color{black}


The model likelihood, based on (\ref{eq:condmodel}) and conditional on the first $L$ observations, is $\prod_{t=L+1}^T f_{Y \mid X}(y_t \mid \bm{y}_{t-1:L}, G)$. This is the form adopted in \citet{antoniano2016} and \citet{kalli2018npbvar}, who assume stationarity, and \citet{deyoreo2017}, who do not assume stationarity. 
The local re-weighting of $\{ \omega_h \}$ with probability density kernels on $\bm{x}$ distinguishes our model from nonparametric extensions of MoE for regression, such as dependent Dirichlet process (DDP; \citealp{maceachern2000ddp}) variants \citep{chung2009probitsb_ssvs, fuentes2009geoweightsddp,barrientos2017} and kernel stick-breaking models \citep{park2010ppm,reich2012varsel}. 
\textcolor{black}{See \citet{wade2014curvefitting} and \citet{MDAKbook} for reviews of density regression models} \color{black} that build on \citet{muller1996curve} \color{black} and \color{black} do not pre-condition the likelihood.

\subsubsection{Covariance factorization}
\label{sec:CholFact}

  To facilitate interpretation in our factorization of the kernels into response and lag densities, allow flexible and parsimonious covariance modeling, and to provide a vehicle for variable selection in the mixture weights, we parameterize the 
\textcolor{black}{Gaussian covariance matrix} according to the factorization
    $\bm{\Sigma} = \bm{B}^{-1}\bm{\Delta}(\bm{B}^{-1})'$. 
  Here, $\bm{\Delta} = \diag (\sigma^2, \delta_{1}^{x}, \ldots, \delta_{L}^{x})$ and $\bm{B}$ is an upper unit-triangular matrix with first row $(1, \beta_{1}^{y},  \beta_{2}^{y}, \ldots, \beta_{L-1}^{y}, \beta_{L}^{y})$, second row $(0, 1, \beta_{1,2}^{x}, \ldots, \beta_{1,L-1}^{x}, \beta_{1,L}^{x})$, and so forth until the $(L-1)$th row $(0, \ldots , 0, 1, \beta_{L-1, L}^{x})$.
  This factorization is equivalent to the square-root-free Cholesky decomposition employed by \citet{daniels2002Cholesky} and \citet{webb2008Cholesky}, and in our setting by \citet{deyoreo2017}. This and similar decompositions have also been used for model selection \citep{smith2002Chol, cai2006cov}. Our extension for lag selection in the mixture weights is discussed in Section \ref{sec:lagselection}.

  This parameterization also yields a sequential decomposition of a joint Gaussian density for $y$ and $\bm{x}$ into $L + 1$ univariate Gaussian densities. Specifically,
  \begin{align}
    \label{eq:sqfChol_seq}
    \Ndist \left( (y, \bm{x} ) \mid \bm{\mu}, \bm{B}^{-1}\bm{\Delta}(\bm{B}^{-1})' \right) 
    = & \Ndist( x_L \mid \mu_L^x, \, \delta_L^x ) \, \times \nonumber \\ 
    &\prod_{\ell = L -1}^1 \Ndist \left( x_\ell \mid \mu_\ell^x - \sum_{r=\ell+1}^L \beta_{\ell,r}^x ( x_r - \mu_r^x ), \delta_{\ell}^x \right) \, \times \nonumber \\
    & \Ndist \left( y \mid \mu^y -  \sum_{\ell = 1}^L \beta_{\ell}^y ( x_{\ell} - \mu_{\ell}^x ), \, \sigma^2 \right) \, . 
  \end{align}
We construct from back (most distant lag) to front ($y$) so that the response density depends on the entire $\bm{x}$ vector while maintaining a consistent order convention. 
%
%
This fully parameterized representation of the covariance matrix is flexible, as each $\beta$ parameter is unrestricted and $\delta$ parameters need only be positive, 
and admits control over the marginal weight density of $\bm{x}$ while preserving positive definiteness. Note also that the marginal covariance matrix of $\bm{x}$ can be constructed as $\bm{\Sigma}^x = (\bm{B}^x)^{-1}\bm{\Delta}^x((\bm{B}^x)^{-1})'$ where $\bm{B}^x$ removes the top row and first column of $\bm{B}$, 
%
%
and $\bm{\Delta}^x = \diag(\delta_1^x, \ldots, \delta_L^x)$.

The weight kernels in $q_h(\bm{x})$ present the most obvious and pressing opportunity to improve parameter economy in the model. We therefore also consider weight kernels with local independence between elements of $\bm{x}$
\citep[e.g.,][]{shahbaba2009}. 
This reduction is accomplished by setting all $\beta^x_{\ell,r}$, for $\ell \ne r$, equal to 0, yielding diagonal $\bm{\Sigma}^x = \bm{\Delta}^x$. We note that Gaussian mixtures with diagonal covariance can approximate 
\textcolor{black}{general density shapes,}
at the cost of possibly utilizing additional mixture components to capture local behavior. The reduction becomes necessary if we include many lags, as the number of covariance parameters for {\em each} component $h$ grows quadratically with $L$.

The final term in (\ref{eq:sqfChol_seq}) involving $\mu^y$ and the $\{ \beta_\ell^y \}$ is overparameterized if used \color{black} as a stand-alone regression model. However, the $\{\mu_\ell^x\}$ parameters become at least partially identified in our mixture formulation because they serve as location parameters for the mixture weight kernels in $q_h(\bm{x})$. \color{black} 
It is nevertheless preferable to monitor inferences for component-specific intercepts $\mu^y + \sum_{\ell=1}^L \beta^y_\ell \, \mu^x_\ell$, which in our experience are far more stable than either $\mu^y$ or $\{ \mu^x_\ell \}$ alone.

%
%
%

\subsubsection{Hierarchical model formulation}
\label{sec:hiermodspec}

\textcolor{black}{
To implement the model, we truncate the infinite summation needed to normalize the mixture weights $\{ q_h(\bm{x}) \}$, using blocked Gibbs sampling \citep{ishwarangibbs2001}.}
There are both theoretical and practical considerations when selecting the truncation level, $H$. Given the DP concentration parameter $\alpha$, we can calculate the prior expected truncation error, 
$\E( \omega_H ) = \E( \prod_{h=1}^{H-1} (1 - v_h) ) = [\alpha/(1+\alpha)]^{H-1} $. We can also monitor throughout MCMC sampling 
the last weight, $\omega_H$, to ensure it remains small, 
as well as the number of occupied components to ensure that it does not approach $H$.

As is common with similar models, we break the mixture by introducing latent variables $\{ s_t \}$ associated with each time point, such that if $s_t = h$, the observation at time $t$ is assigned to component $h$. We denote all component-specific parameters as $\{ \bm{\eta}_h \}_{h=1}^H$ where $\bm{\eta} \equiv \{ \mu^y, \bm{\mu}^x, \bm{\beta}^{y}, \bm{\beta}_1^x, \ldots, \bm{\beta}_{L-2}^x, \beta_{L-1}^x, \sigma^2, \bm{\delta}^x \}$, with vectors $\bm{\beta}^y$ and $\bm{\beta}_\ell^x$ (for $\ell = 1, \ldots, L-2$), and $\beta_{L-1}^x \equiv \beta_{L-1, L}^x$ taken from the corresponding rows of $\bm{B}$, and $\bm{\delta}^x = (\delta_1^x, \ldots, \delta_L^x)$. 
Again, we use notation $\Ndist_{(h)}(\cdot)$ to indicate that all parameters used to specify the mean and covariance are indexed by $h$. The hierarchical formulation of our model is given by
%
%
  \begin{align}
    \label{eq:hierarchicalmodel}
    \begin{split}
    y_t \mid \bm{y}_{t-1:L}, s_t = h, \{\bm{\eta} \} &\simindep \Ndist_{(h)} \left( y_t \mid \mu^y - \sum_{\ell=1}^L \beta_{\ell}^{y} ( y_{t-\ell} - \mu_{\ell}^x ), \, \sigma^2 \right),  \\ & \quad \text{for} \ t = L+1, \ldots, T, \ \text{and} \ h = 1, \ldots, H,  \\
    \Pr(s_t = h \mid \bm{y}_{t-1:L}, \{\bm{\eta}\}, \bm{\omega}) &= \frac{ \omega_h \Ndist_{(h)}( \bm{y}_{t-1:L} \mid \bm{\mu}^x, \bm{\Sigma}^x) }{ \sum_{j=1}^H \omega_j \Ndist_{(j)}( \bm{y}_{t-1:L} \mid \bm{\mu}^x, \bm{\Sigma}^x) } \, , \\ 
    \omega_1 = v_1, \ \omega_h = v_h \prod_{j=1}^{h-1} &(1 - v_j), \ \text{for}\ j=2,\ldots,H-1, \ \text{and } \omega_H = \prod_{j=1}^{H-1} (1 - v_j) \, , \\ 
    v_j \mid \alpha &\simiid \Betadist(1, \alpha), \ \text{for} \ j = 1, \ldots, H-1, \\ 
    \bm{\eta}_h \mid G_0 &\simiid G_0(\bm{\eta}_h), \ \text{for} \ h = 1, \ldots, H, 
    \end{split}
  \end{align}
with $G_0( \bm{\eta} ) = \Ndist((\mu^y, \bm{\beta}^y) \mid \sigma^2) \times \IGdist(\sigma^2) \times \Ndist(\bm{\mu}^x) \times \prod_{r=1}^{L-1} \Ndist(\bm{\beta}_r^x) \times \prod_{\ell=1}^L\IGdist(\delta_\ell^x)$, and $\bm{\omega} = (\omega_1, \ldots, \omega_H)$. Here $\Ndist((\mu^y, \bm{\beta}^y) \mid \sigma^2)$ indicates that the prior covariance matrix for $\bm{\beta}^* \equiv (\mu^y, \bm{\beta}^y)$ is scaled by $\sigma^2$, which allows us to analytically integrate all $y$-indexed parameters from the full conditional for $\bm{\eta}_h$ and improve mixing in MCMC (discussed in Section \ref{sec:computation}).

  We complete the model with 
\textcolor{black}{a $\Gammadist (a_\alpha, b_\alpha)$ prior for $\alpha$, 
and with}
  conditionally conjugate priors on the parameters in $G_0$. Specifically, the $(L+1)$-variate Gaussian distribution for $\bm{\beta}^{*}$ has mean $\bm{\beta}_0^{*} \sim \Ndist(\bm{b}_0^*, \bm{S}_0^*)$ and covariance $\sigma^2 (\bm{\Lambda}_0^*)^{-1}$ with $(\bm{\Lambda}_0^*)^{-1} \sim \IWishdist(\nu^*, \nu^*\bm{\Psi}_0^*)$ (an inverse-Wishart distribution with $\nu^*$ degrees of freedom and mean $\nu^* \bm{\Psi}_0^*/[\nu^* - (L+1) - 1]$, parameterized so that $\bm{\Psi}_0^*$ is the prior harmonic mean of $(\bm{\Lambda}_0^*)^{-1}$). The inverse-gamma distribution for $\sigma^2$ has fixed shape $\nu_{\sigma^2} / 2$ and scale $\nu_{\sigma^2} \, s_0 / 2$, yielding for $\sigma^2$ a prior harmonic mean of $s_0 \sim \Gammadist(a_{s_0}, b_{s_0})$ (which itself has mean $a_{s_0}/b_{s_0}$). The Gaussian distribution for $\bm{\mu}^x$ has mean $\bm{\mu}_0^x \sim \Ndist(\bm{m}_0^x, \bm{S}_0^{\mu_x})$ and covariance $(\bm{\Lambda}^{\mu_x})^{-1} \sim \IWishdist(\nu^{\mu_x}, \nu^{\mu_x} \bm{\Psi}_0^{\mu_x})$. The Gaussian distribution for each $\bm{\beta}_r^x$ has mean $\bm{\beta}_{0,r}^x \simindep \Ndist(\bm{b}_{0,r}^{\beta_x}, \bm{S}_{0,r}^{\beta_x})$ and covariance $(\bm{\Lambda}_{0,r}^{\beta_x})^{-1} \simindep \IWishdist(\nu_r^{\beta_x}, \nu_r^{\beta_x} \bm{\Psi}_{0,r}^{\beta_x})$, for $r=1,\ldots,L-1$. The inverse-gamma distribution for each $\delta_\ell^x$ has fixed shape $\nu_\ell^{\delta^x} / 2$ and scale $\nu_\ell^{\delta^x}  s_{0,\ell}^x / 2$ with $s_{0,\ell}^x \simindep \Gammadist(a_{s_{0},\ell}^x, b_{s_0,\ell}^x)$, for $\ell=1,\ldots,L$.

\color{black}
Experience with the model suggests it is practical to fix components in $G_0$ associated with $y$-indexed parameters rather than use the full prior specification above. Specifically, we find that fixing $\bm{\beta}_0^*$ at $\bm{b}_0^*$, $\bm{\Lambda}_0^*$ at $(\bm{\Psi}_0^*)^{-1}$, and $s_0$ at a prior guess $s_{00}$ works well in practice.
\color{black}

\subsection{Prior settings}
\label{sec:prior_settings}


  The priors for the hierarchical model in Section \ref{sec:hiermodspec} are specified in generality so that the model can be fit with the time series $\{ y_t \}$ at any scale and for a variety of functional characteristics. However, one may consider first removing certain known trend and cyclical behaviors, and basing hyperparameter settings on default values. Here, we recommend default values derived from marginal summaries of the time-series.

\textcolor{black}{We first discuss the function and interpretation of model parameters. A key} consideration is that model (\ref{eq:condmodel}) is a locally weighted mixture of Gaussian linear regression models. The weight structure depends not only on $\{ \omega_h \}$, which is inherited from the nonparametric prior and (for low values of $\alpha$) encourages economy in clustering, but also on the Gaussian kernels for $\bm{x}$. One could imagine a normalized weight 
  surface spanning $\mathbb{R}^L$ for each mixture component $h$ that follows the contours of a $L$-variate Gaussian density weighted by $\omega_h$. The component-specific, $x$-indexed parameters, $\bm{\mu}^x$ and 
  $\bm{\Sigma}^x =$ $(\bm{B}^x)^{-1}\bm{\Delta}^x((\bm{B}^x)^{-1})'$, determine the locations and shapes of the weight kernels. The $y$-indexed parameters, $\mu^y$ and $\bm{\beta}^y$, provide the component-conditional mean as a first-order linear combination of $\bm{x}$, and $\sigma^2$ provides observation error variance around the component's mean.

  One primary functional of interest derived from the transition density in (\ref{eq:condmodel}) is the conditional expectation $\E(y \mid \bm{x} )=$
  $\sum_{h} q_h(\bm{x}) \, \mu_{h}(\bm{x})$, to which we refer as the transition mean. A modeler can encode beliefs about this functional relationship between $y$ and $\bm{x}$ through the priors for $\alpha$ and parameters in the base measures for $\bm{\Sigma}^x$ and $\sigma^2$. By influencing the number of \color{black} occupied mixture components in this locally linear model, $\alpha$ assists in controlling complexity of the global transition mean. \color{black} 
 To encourage smooth behavior, one may use a prior favoring relatively large variances in $\bm{\Sigma}^x$, most directly through the priors for $\{\delta^x_\ell\}$. To encourage active local behavior, including nearly discontinuous transitions, one would use small variances in $\bm{\Sigma}^x$ to allow the components to concentrate on small regions, analogous to using many knots in spline models. 
Supplement \ref{sec:appendix_priorsettings} further explores the effect of prior settings on transition means.

  We recommend the following default settings for a baseline prior, which in most cases should be adjusted for the analysis at hand. We typically set $a_\alpha$ in the interval $[5,15]$, depending on our prior beliefs about the degree of nonlinearity in the transition function. Setting $b_\alpha=1$ yields a prior mean of $a_\alpha$. \citet{antoniak1974mixtures} gives the expression $\alpha \log\left( (\alpha + T - L) / \alpha \right)$ as a rough prior estimate for the number of components. While this applies in the prior joint model, the number of components in our conditional model (\ref{eq:condmodel}) is also a function of the Gaussian weight kernels on $\bm{x}$. We set $\bm{b}_0^* = (\bar{y}, 0, \ldots, 0 )$, with $\bar{y}$ representing the center of the time series\color{black}, either empirical or based on prior information, thus centering the model.
%
We use $\bm{\Psi}_0^* = s_{00}^{-1} \diag( [\range(y)/2.0]^2, 16.0, \ldots, 16.0)$, where $s_{00}$ is a user-supplied prior guess of $\sigma^{2}$, and $\range(y)$ represents the range of the time series, either empirical or based on prior information. The prior guess $s_{00}$ partially compensates and controls for the fact that the covariance for $\bm{\beta}^*$ in $G_0$ is multiplied by $\sigma^2$. 
We use $s_{00} = [\range(y) / 6.0 ]^2 / \mathcal{R}$ as an automatic prior guess of $s_0$. \color{black} The squared quantity is divided by a prior signal-to-noise ratio $\mathcal{R} > 0$ that is set by the modeler on a case-by-case basis. 
\color{black} We interpret $\mathcal{R}$ roughly as the ratio of total variance to mixture-component error variance. \color{black}
We typically use $\mathcal{R} \in [5.0, 10.0]$. We use $\bm{m}_0^x = \bar{y} \, \bm{1}$ and $\bm{S}_0^{\mu_x} = [\range(y) / 6.0]^2 \, \bm{I}_{L}$, where $\bm{I}_{k}$ denotes a $k \times k$ identity matrix. We allow for variability in $\bm{\mu}^x$ by setting $\nu^{\mu_x} = 10 \, (L + 2) $ and $\bm{\Psi}_0^{\mu_x} = [ \range(y) / 2.0]^2 \, \bm{I}_L$. Similarly, we set each $\bm{b}_{0,r}^{\beta_x} = \bm{0}$, each $\bm{S}_{0,r}^{\beta_x} = \bm{I}_{L}$, each $\nu_r^{\beta_x} = 10 \, (L+2)$ and $\bm{\Psi}_{0,r}^{\beta_x} = 2.0 \, \bm{I}_{L-1-r+1}$, for $r = 1, \ldots, L-1$. Finally, we set $\nu_\ell^{\delta_x} = 5.0$, with $a_{s_0,\ell}^x = n_{s_0,\ell}^x \, \nu_\ell^{\delta_x} / 2$ and $b_{s_0,\ell}^x = n_{s_0,\ell}^x \, \nu_\ell^{\delta_x} / (2 \, s_{00, \ell}^x) $, for $\ell = 1, \ldots, L$, where $n_{s_0, \ell}^x = 5.0$ and $s_{00, \ell}^x = [ \range(y) / 8.0 ]^2$.

  While the preceding prior settings provide a good starting point in general, they are not always appropriate. We recommend considering alternate settings, especially for $\alpha$, and parameters in the base measures for $\bm{\Sigma}^x$ and $\sigma^2$, depending on prior beliefs about the functional relationship being modeled in each analysis. We further recommend checking for sensitivity of inferences for important quantities to these and other prior settings. \color{black} Supplement \ref{sec:appendix_sensitivity} reports a simulation study exploring sensitivity of posterior inference results to changes in $\alpha$ and $\mathcal{R}$.
\color{black}

\subsection{Computation}
  \label{sec:computation}

  We briefly outline the MCMC algorithm used to obtain posterior samples from the proposed model. Further details are given in 
Supplement \ref{sec:appendix_MCMC}. 
We employ a Gibbs sampler with a variety of update methods for parameter blocks, which proceeds by successively sampling the parameters in the sets and manner described below.

  \noindent{\bf Latent states}: 
  The latent states identifying component membership for each observation $y_t$ are updated individually, each using a Metropolized Gibbs step \citep{liu1996gibbs} based on discrete full conditional distributions involving $\{ \omega_h \}$, the weight kernel density 
\textcolor{black}{for $\bm{y}_{t-1:L}$,} and the kernel density for $y_t$.

  \noindent{\bf Stick-breaking weights}:
\textcolor{black}{The DP weights $\{ \omega_h \}_{h=1}^H$}  
  are defined through the latent $\{ v_h \}_{h=1}^{H-1} $ which, conditional on component membership $\{ s_t \}$, admit $H-1$ independent beta full conditional distributions in standard DPM models \citep{ishwarangibbs2001}. The normalization term in each likelihood contribution of 
  $q_h(\bm{y}_{t-1:L})$
  complicates the full conditional distribution in our model. It is unchanged from the distribution reported in \citet{deyoreo2017}, with the exception that the kernels are now multivariate Gaussian on the vector 
  $\bm{y}_{t-1:L}$. This 
  adjustment yields numerical instability and poor mixing in the one-at-a-time slice sampler employed by \citet{deyoreo2017}. To obtain direct samples from this distribution, we instead employ the multivariate hyper-rectangle slice sampler of \citet{neal2003slice} to update all $v_h$, $h = 1, \ldots, H-1$, simultaneously.

  \noindent{\bf Component-specific parameters}:
  To facilitate mixing of the $y$-indexed, component-specific parameters, we partition $\bm{\eta}$ into its $y$ and $x$ components $\bm{\eta}^y \equiv \{ \mu^y, \bm{\beta}^y, \sigma^2 \}$ and $\bm{\eta}^x \equiv \{ \bm{\mu}^x, \bm{\beta}_1^x, \ldots, \beta_{L-1}^x, \bm{\delta}^x \}$, and sample $p(\bm{\eta}_h \mid \cdots) = p(\bm{\eta}^x_h \mid \cdots, -\bm{\eta}^y_h) \ p( \bm{\eta}^y_h \mid \bm{\eta}^x_h, \cdots)$, where $p(\bm{\eta}^x_h \mid \cdots, -\bm{\eta}^y_h) = \int p(\bm{\eta}_h \mid \cdots) \, \diff \bm{\eta}^y_h$. The weight normalization terms in 
  $q_h(\bm{y}_{t-1:L})$ preclude simple conjugate updates of $\bm{\eta}_h^x$, for which we employ a random-walk Metropolis 
\textcolor{black}{step. This is then} followed by an exact draw from the full conditional distribution of $\bm{\eta}_h^y$.

\noindent{\bf DP prior hyperparameters}:
  All parameters of the DP centering distribution have 
  conditionally conjugate updates. For computational stability, our implementation fixes, rather than updates, the parameters in $G_0$ associated with $\bm{\eta}^y$ \color{black} at prior summary values, as noted in Section \ref{sec:hiermodspec}. 
  %
%
\color{black}
\textcolor{black}{Finally, the DP concentration parameter $\alpha$ has a gamma posterior full conditional distribution with shape $a_\alpha + H - 1$ and rate $b_\alpha - \log(\omega_H)$.}  

%
%


  We typically initialize MCMC chains at default prior settings such as the prior mean or applicable summary value from the next level of the hierarchy, or with draws from the prior model (usually with $G_0$ fixed). The primary exception is the initial allocation to components $\{ s_t \}$, for which we use output from a clustering algorithm applied to 
\textcolor{black}{$(y_{t},y_{t-1},...,y_{t-L})$,}  
  for all $t = L+1, \ldots, T$. For example, we use hierarchical clustering with Euclidean distance and Ward linkage to assign the observations into $H$ clusters. The sampler is then run for one or several rounds of tuning or adaptation, as described in Supplement \ref{sec:appendix_MCMC}. 
  If adaptation is used, scaled empirical covariance matrices inform subsequent random-walk proposals. 
  After a specified burn-in period, samples are collected for inference.

  In our experience, the weakly identified $\bm{\eta}^x$ and $\bm{\omega}$ parameters present the primary mixing challenge. This appears to indicate redundancy in the weight functions, for which many configurations produce similar results. Our illustrations with the base model 
\textcolor{black}{(i.e., without lag selection)}  
  focus on low-order dependence $L \le 5$. Later illustrations use diagonal $\bm{\Sigma}^x = \bm{\Delta}^x$\color{black}, which reduces the computational complexity of the most expensive update, for $\{ \bm{\eta}^x_h \}$, from ${O}(THL^3)$ to ${O}(TL^2 + HL^3 + THL)$\color{black}. We further aid mixing by iterating between adaptation and pre-burn-in runs before beginning a final burn-in run. We note that despite the mixing challenges, \color{black} estimates for functionals of interest \color{black} are typically stable.
  
 \color{black}
  MCMC and other computations for the proposed model were run in the {\it Julia} language \citep{bezanson2017julia}. Runtimes under various settings are compared as part of a sensitivity analysis in Supplement \ref{sec:appendix_sensitivity}.
\color{black}

  \subsection{Transition density estimation}
  \label{sec:transdens_estimation}
  
  Posterior samples from the model yield rich inferences regarding the transition distribution for a time series. The three of most interest to us are the transition density, the transition mean functional, and inferences for relevant lags. We incorporate the latter in Section \ref{sec:lagselection}. The transition mean functional and estimates of the transition density 
  are straightforward to compute, as the stick-breaking representation and blocked Gibbs sampler yield 
  an approximation of the random mixing distribution $G$ at each MCMC iteration. For any value of $y$ and $\bm{x}$, or over a multidimensional grid of values, one can use posterior samples of parameters to calculate pointwise samples of the finite-truncated version of $f_{Y|X}$ in (\ref{eq:condmodel}), given as 
   $ \tilde{f}_{Y|X}(y\mid\bm{x}) =  \sum_{h=1}^H \tilde{q}_h(\bm{x} ) \, \Ndist_{(h)}(y\mid \mu(\bm{x}), \sigma^2 )$, 
  with $\tilde{q}_h(\bm{x}) = \omega_h \Ndist_{(h)}(\bm{x}) / \sum_{j=1}^H \omega_j \Ndist_{(j)}(\bm{x})$ and $\mu(\bm{x}) = \mu^y - \sum_{\ell=1}^L \beta^y_\ell \, (x_\ell - \mu_\ell^x)$. The samples can then be used to 
\textcolor{black}{construct point and interval estimates for the transition density.}  
  Other functionals such as the transition mean or quantiles are similarly obtained. One can calculate the transition mean for each posterior sample with $\tilde{\E}_{Y|X}(y\mid\bm{x}) = \sum_{h=1}^H \tilde{q}_h(\bm{x}) \, \mu_{(h)}(\bm{x} )$ over a grid of values for $\bm{x}$, yielding pointwise estimates and intervals. We obtain samples of the $u \in (0,1)$ quantile of the transition density by solving for the unique root of
   $ \tilde{Q}_u (y \mid \bm{x}) = u - \sum_{h=1}^H \tilde{q}_h(\bm{x}) \, \Phi\left([y - \mu_{(h)}(\bm{x})] /  \sigma_{(h)} \right) $, 
  where $\Phi(\cdot)$ is the standard normal cumulative distribution function.
  
  Monte Carlo estimates of $K$-step-ahead forecasts can be obtained by inductively simulating $(s, y)_{T+k}$ pairs, for $k = 1, \ldots, K$, following the first two levels of the hierarchical model (\ref{eq:hierarchicalmodel}) for each posterior sample. Such samples propagate both forecast and inferential uncertainty, and can be useful for assessing model performance with validation data.

\section{Data illustrations}
\label{sec:illustrations}
  
  We illustrate the proposed model with two examples. The first synthetic data example highlights some key features and potential uses of the model. The real data example illustrates the model's utility for lag-dependent density estimation. Two default prior settings were utilized in each case, with one promoting a higher signal variance through prior signal-to-noise ratio $\mathcal{R} = 8.0$ instead of the default $\mathcal{R} = 5.0$. For each model fit, multiple MCMC chains were randomly initialized using the strategy described in Section \ref{sec:computation}, followed by iterative tuning (no adaptation) and 300,000 burn-in samples. The next 500,000 iterations were then thinned to 5,000 for inference (plots in the following illustrations generally use 1,000 or 2,000 of these). 
  Inferences are reported for one of the chains. 
 These values for burn-in and thinning are fairly conservative; shorter chains often suffice.

  \subsection{Simulated data: Ricker model}
  \label{sec:ex_appdataLag2}
  
\begin{figure}[t!]
    \includegraphics[width=2.25in]{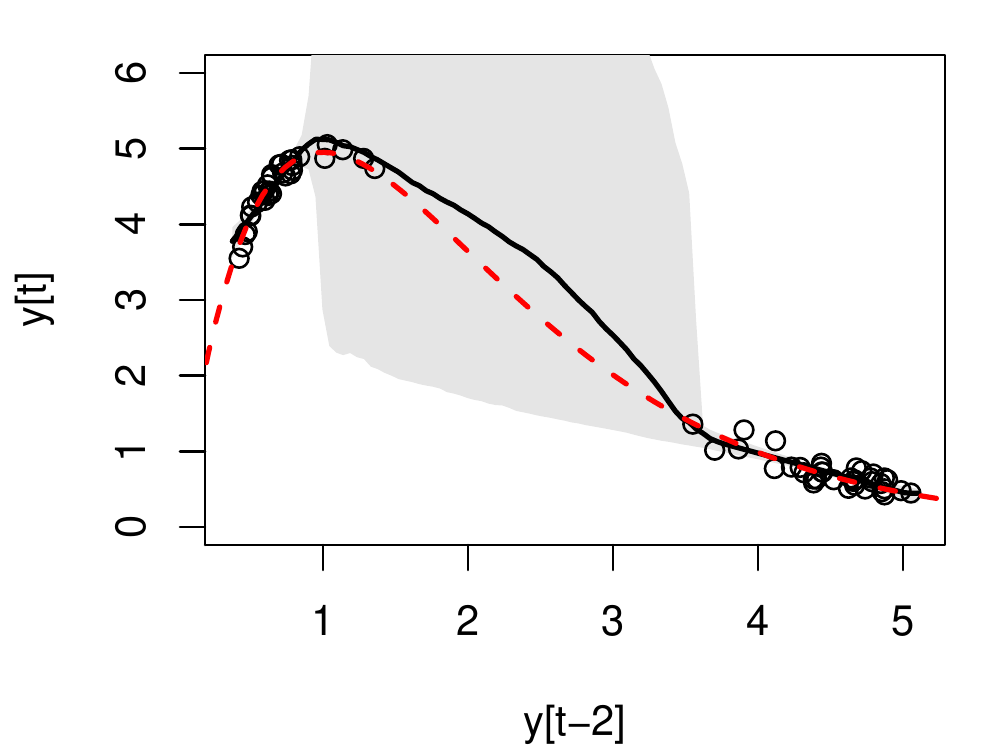}       
    \includegraphics[width=2.25in]{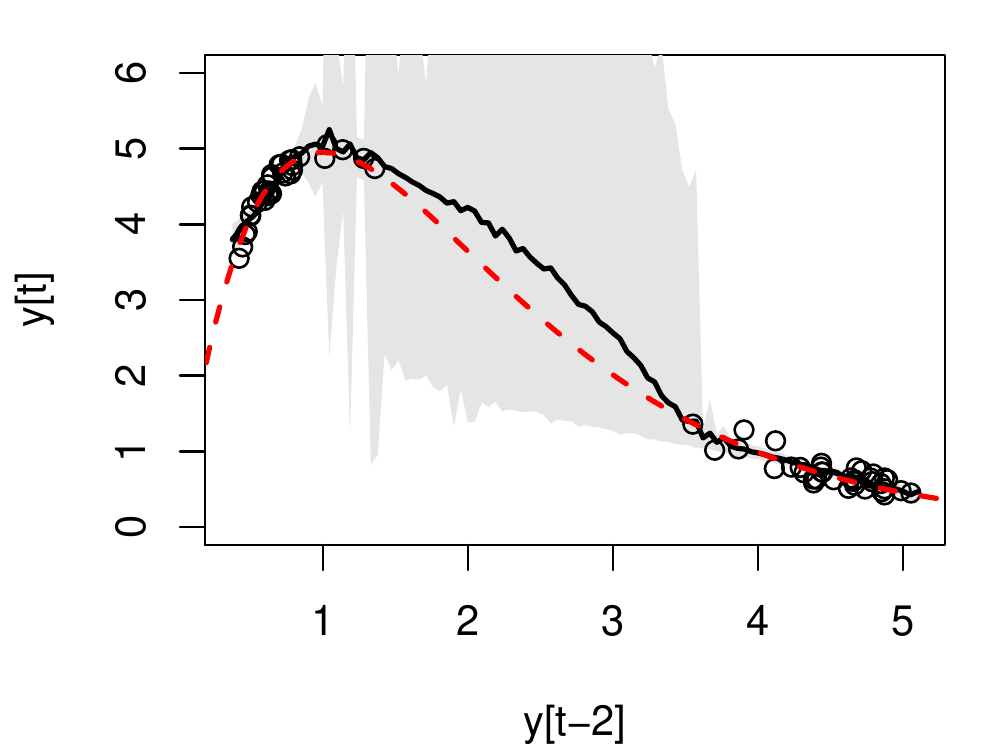}
  \caption{{\small Model fit to the single-lag dynamical simulation with noise ($T=72$, $L=2$, $\mathcal{R}=8.0$), 
  depicting the posterior mean and 95\% interval estimates for the transition mean function over a grid of values for lag 2. In the left panel, values of the first lag ($y_{t-1}$) were fixed at a mean value. In the right panel, values for the first lag were randomly drawn uniformly over the range of observed values. Observations are included, as well as the true transition map (dashed red line).}}
  \label{fig:appdata_lag2only_transMean}
\end{figure}

  We begin with a time series simulated from an adaptation of a classical model for population dynamics \citep{ricker1954}. The series was generated from
  \begin{align}
    \label{eq:appDatalag2gen}
    y_t = y_{t-2} \exp(2.6 - y_{t-2}) + \epsilon_t \, , \quad \epsilon_t \simiid \Ndist(0, (0.09)^2) \, ,
  \end{align}
  featuring first-order nonlinear dynamics as a function of the second lag only. We fit the 
  model to the original real-valued time series with $L=2$, $T=72$ (so that 70 observations contribute to the likelihood), and $H=40$. The $\mathcal{R}=5.0$ fit resulted in three chains with similar traces of the log-likelihood and occupied mixture components (always at two). All traces of $\sigma^2$ for the most occupied cluster (not shown) converge to approximately 3.5 times the true value of 0.0081, due in part to the prior estimate $s_{00}=0.119$. 
  Flexibility and prior bias in error variance, together with low sample size, result in a transition mean fit that locally mixes two planes, capturing the general shape, but missing curvature in the region $y_{t-2} \in (0,1)$ (not shown). 
  Two of three chains with higher signal-to-noise ratio ($\mathcal{R}=8.0$) use a third mixture component to better capture this curvature (although one reverts back to two components), as demonstrated for one chain in Figure \ref{fig:appdata_lag2only_transMean}.

  The dynamics are reasonably recovered in data-rich regions of the phase space despite using an over-specified model with two lags on a short time series. 
%
  We can informally assess the influence of the first lag with the second-order model by checking for sensitivity of inferences for the transition mean to different values of the first lag. For example, the left panel of Figure \ref{fig:appdata_lag2only_transMean} plots 
  estimates for the transition mean over a grid of values for the second lag, in which all values for the first lag have been fixed at their mean. The right panel replicates this plot with grid values for the first lag drawn uniformly over the range of the data. This perturbation has minimal effect, especially where data are observed, suggesting that lag 1 is negligible in the model fit. We note particularly wide credible intervals in the data-sparse region, which approximately reach 10. This appears to stem from the weight functions concentrating locally around the data, leaving data-sparse regions to revert to an indecisive mixture of the component fits and prior.

  \subsection{Old Faithful data}
  \label{sec:ex_oldFaithful}

\begin{figure}[t!]
      \includegraphics[scale=0.5, trim = 0 10 0 40, clip]{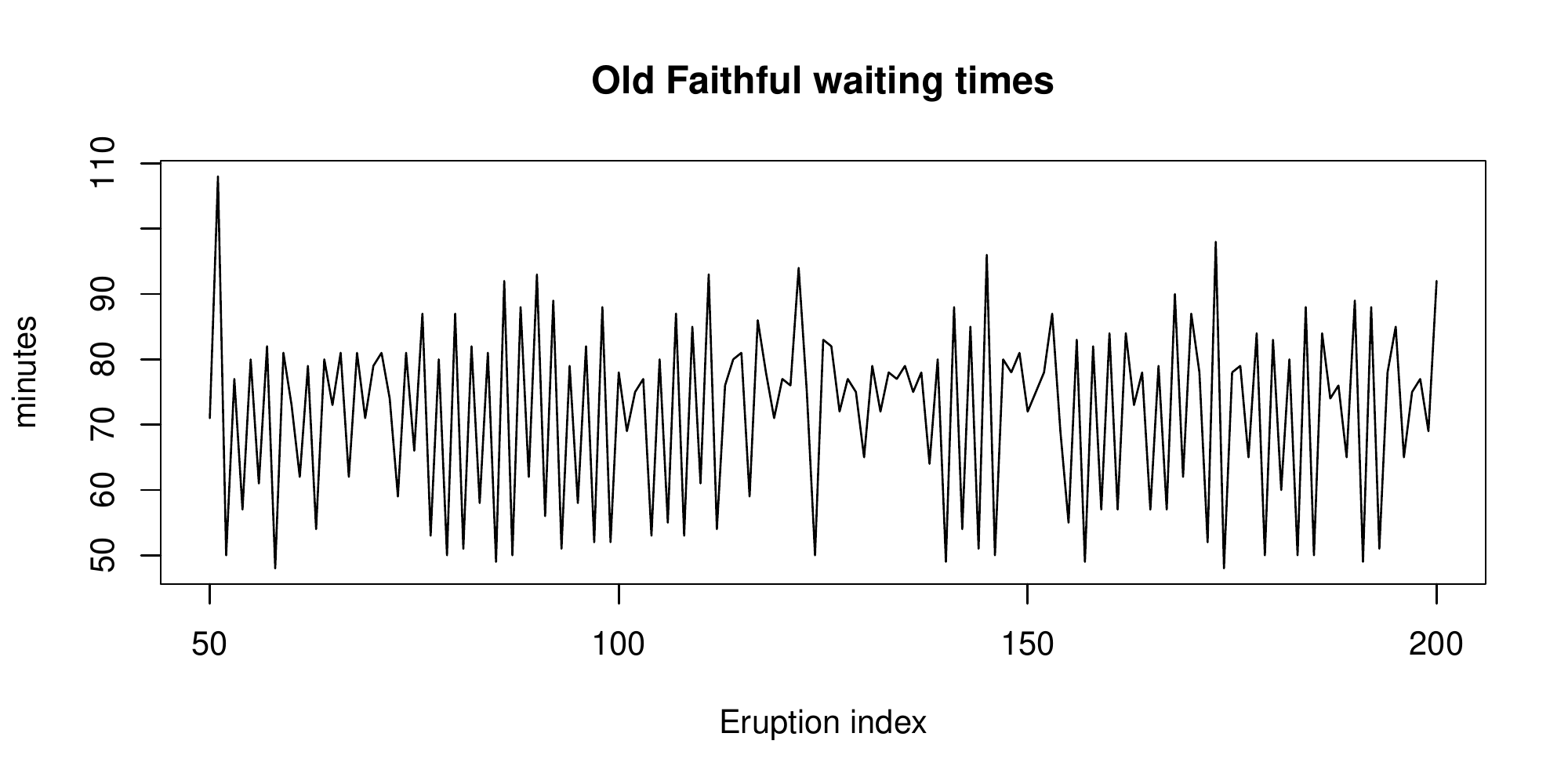}
    \caption{{\small Trace of 150 consecutive Old Faithful eruption waiting times in minutes. This window of the middle half of the time series typifies the data, with exception of the run of long waiting times between index 120 and 140.}}
    \label{fig:OldFaithful_ts}
\end{figure}

  \citet{antoniano2016} and \citet{deyoreo2017} both illustrate single-lag versions of our proposed model with the well-known inter-eruption waiting times of the Old Faithful geyser in Yellowstone National Park, U.S.A. The time series has attracted attention, both for illustration and analysis from chaos \citep{nicholl1994} and statistical \citep{azzalini1990} perspectives, partly due to nonlinear as well as non-Gaussian dynamics. We revisit Old Faithful using the traditional data set reported in \citet{azzalini1990}, consisting of 299 consecutive pairs of eruption durations and waiting times between August 1 and 15, 1985. Figure \ref{fig:OldFaithful_ts} shows a trace of eruption waiting times in minutes. 
  
\begin{figure}[t!]
    \includegraphics[height=1.8in]{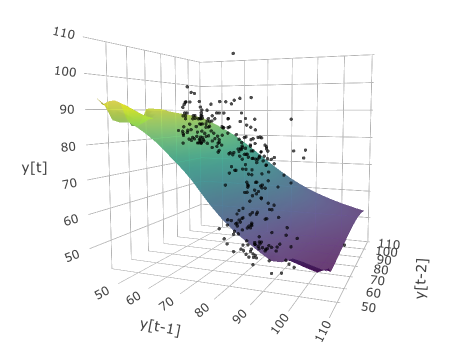}
    \includegraphics[height=1.8in]{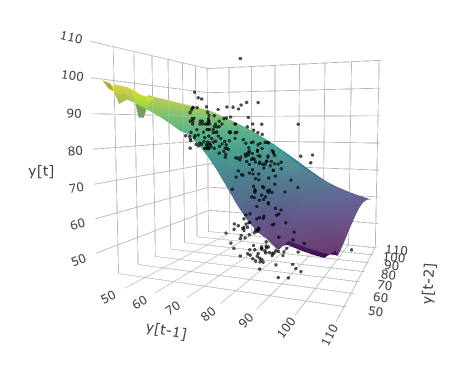}
  \caption{{\small Nonparametric model fit to Old Faithful waiting times in minutes ($T=291$, $L=2$, $\mathcal{R}=5.0$), with 
  posterior mean estimates of the transition mean (left) and 0.8 quantile (right) surfaces. Observed transitions are included as points.}}
  \label{fig:OldFaithful_transMean}
\end{figure}

  We fit the proposed model to the final $T=291$ observations with $L=2$ and $H=40$. Likelihood traces are similar among runs under both prior signal-to-noise ratios, switching (infrequently) between values corresponding to two and three occupied mixture components. Estimated transition mean surfaces, one of which is shown in Figure \ref{fig:OldFaithful_transMean} (left), are primarily driven by the first lag, with minor tilt along the second. The transition mean functional is less informative for values of $y_{t-1}$ above 70 minutes, when the transition distribution becomes bimodal. In this region, estimates of transition quantiles may be more appropriate than the transition mean. Inferences for quantiles over a grid of fixed lag values are easily obtained from posterior samples by following the procedure described in Section \ref{sec:transdens_estimation}. Figure \ref{fig:OldFaithful_transMean} (right) shows a pointwise posterior mean estimate of the 0.8 quantile surface as a function of the two lags. Credible intervals for both surfaces (excluded for simplicity in the plots) are reasonable, falling within the range of the data.    
 
\begin{figure}[t!]
    \includegraphics[height=1.65in, page=2, trim=5 0 15 0, clip]{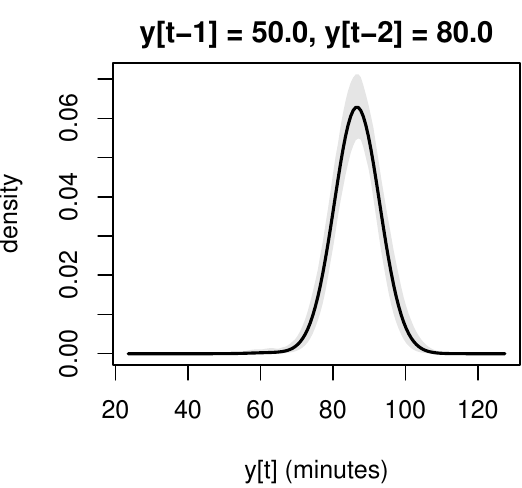}
    \hspace{-5pt}
    \includegraphics[height=1.65in, page=3, trim=24 0 15 0, clip]{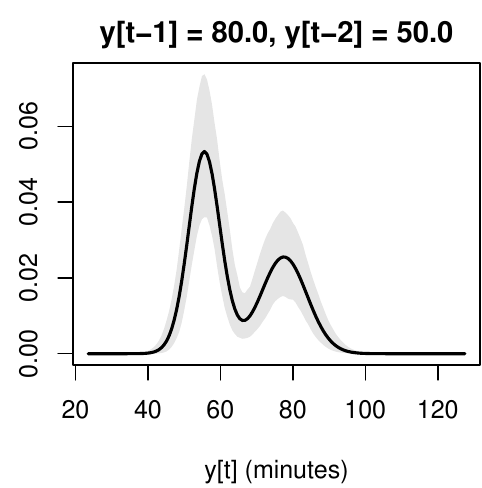}
    \hspace{-5pt}
    \includegraphics[height=1.65in, page=4, trim=24 0 15 0, clip]{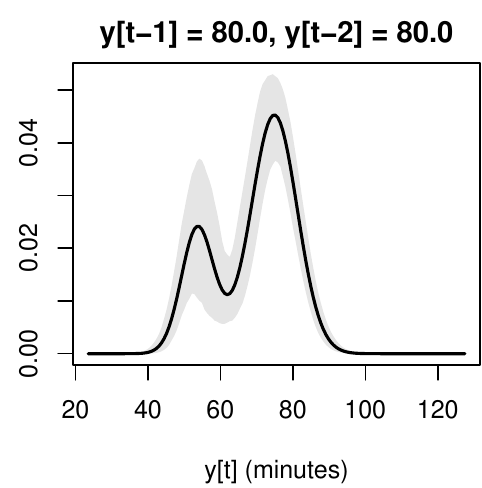}
  \caption{{\small Posterior mean and 95\% interval estimates for the transition density of Old Faithful waiting times at three pairs of fixed values of the first two lags ($T=291$, $L=2$, $\mathcal{R}=5.0$).}}
  \label{fig:OldFaithful_densities}
\end{figure}

Figure \ref{fig:OldFaithful_densities} shows estimated transition densities (posterior mean and 95\% credible intervals) for three values of the two lags. These estimates demonstrate the density autoregressive feature of the model, which in this case successfully captures density dependence on lags. Interestingly, the transition density undergoes noticeable change between $y_{t-2} = 50$ and $y_{t-2} = 80$ when $y_{t-1}$ is fixed at 80 minutes, suggesting 
dependence on the second lag.
Other runs show similar structure. 
A simple analysis using a discrete-state Markov chain on a dichotomized version of the time series further 
supports second-order dependence.

\section{Lag selection} 
\label{sec:lagselection}

We now discuss extending model (\ref{eq:condmodel}) to include inferences for relevant lags. This step is important in many applications, as dependence may extend beyond the most recent lags. In some cases, not all recent lags are important. Methods for state-space reconstruction require a minimal number of lags to ``unfold'' an attractor, but using too many can be inefficient, or render estimation impractical. Reducing system dimensionality to the minimum necessary for fitting the data further simplifies posterior analysis and model interpretation. Our approach is to pre-specify a maximal lag horizon $L$, and fit an encompassing model that accommodates up to all $L$ lags, but shrinks to select only those that significantly contribute to the transition density.

In the time series literature, autoregressive order is often assessed with standard information criteria, which can include regularization \citep{khalili2017regularization}. Bayesian approaches typically involve stochastic-search-type algorithms, and several are presented in \citet[Ch.\ 2]{prado2010}. In the stationary, linear case, one can use the specialized priors of \citet{huerta1999priors} on roots of the AR characteristic polynomial to infer order. \citet{wood2011mixAR} employ a two-stage MCMC sampler on a time-weighted mixture of autoregressive models to infer component-specific order and perform Bayesian model averaging. 

\citet{ohara2009review} provide a review of Bayesian variable selection methods in the regression setting, including that of \cite{kuo1998variableselect}, which we adopt here. 
There is also a growing literature for variable selection in BNP regression modeling. \citet{barcella2017review} provide a 
review that discusses approaches for covariate-dependent DPM, DDP, and product partition models. Most approaches involve binary indicator variables associated with each covariate that either activate lag-specific kernels (as in \citealp{reich2012varsel}) or break mixtures for key parameters (i.e., regression coefficients) involving point masses at 0 (as in \citealp{chung2009probitsb_ssvs}). Another option with DPM models is to include model order as a mixing parameter (as in \citealp{lau2008dpmar}).

We propose a model extension for global lag selection in Section \ref{sec:lagselection_extensions}. Section \ref{sec:inference_lagselect} discusses inference, including posterior sampling and other modifications to MCMC, and sampling for functionals. Section \ref{sec:local_lagsel} describes an analogous extension for local lag selection. In Section \ref{sec:illustrations_lagselect}, we revisit the data illustrations from Section \ref{sec:illustrations} and include two additional data sets.

\subsection{Model extension \textcolor{black}{for global lag selection}}
\label{sec:lagselection_extensions}

In model (\ref{eq:condmodel}), both mixture kernels and weights depend on 
\textcolor{black}{the lags, thus} necessitating coordination across multiple parameters for model-based 
\textcolor{black}{lag selection}.
To this end, we employ binary variables $\{ \gamma_\ell \} $, for $\ell=1,\ldots,L$, to indicate dependence of 
\textcolor{black}{$y_{t}$ on $y_{t-\ell}$}, \color{black} in both weights and kernels of all $H$ mixture components, \color{black} 
if $\gamma_\ell=1$. The most straightforward approach to incorporating these indicators follows \citet{kuo1998variableselect}, wherein we replace $\beta^y_\ell$ with $\gamma_\ell \, \beta^y_\ell$. 
The modification to $\bm{\beta}^y$ controls lag dependence in the mixture kernels. 
Our proposed modification to the weight kernels 
$\Ndist_{(h)}(\bm{y}_{t-1:L} \mid \bm{\mu}^x, \bm{\Sigma}^x)$ 
totally eliminates dependence on lags \color{black} for which $\gamma_\ell = 0$\color{black}, and is most clearly understood in the context of the sequential construction of weight kernels given in (\ref{eq:sqfChol_seq}). We replace $\beta^x_{\ell,r}$ with $\gamma_\ell \, \gamma_r \, \beta^x_{\ell,r}$. \color{black} Additionally, \color{black} if $ \gamma_\ell = 0$, the univariate Gaussian density associated 
\textcolor{black}{with $y_{t-\ell}$}
is replaced with $1$. This is equivalent to appropriately subsetting 
$\{ \bm{\beta}_\ell^x \}$ and $\bm{\delta}^x$ prior to constructing the covariance matrix $\bm{\Sigma}^x$, reducing the dimensionality of 
$\Ndist_{(h)}(\bm{y}_{t-1:L} \mid \bm{\mu}^x, \bm{\Sigma}^x)$ 
to $n_\gamma = \sum_{\ell=1}^L \gamma_\ell$. If $n_\gamma = 0$, then the weight function reduces exclusively to $\bm{\omega}$, resulting in a 
\textcolor{black}{standard univariate Gaussian DPM model.}
This approach reduces computational burden and offers a clean, complete lag selection, conditional on $\bm{\gamma} = (\gamma_1, \ldots, \gamma_L)$.


The modification for lag selection affects the hierarchical model in (\ref{eq:hierarchicalmodel}) 
through 1) the regression mean in the mixture kernel distribution for $y_t$, which becomes $\mu^y - \sum_{\ell = 1}^L \gamma_\ell \, \beta_\ell^y \, (y_{t-\ell} - \mu_\ell^x)$; 2) the construction of $\Ndist_{(h)}(\bm{y}_{t-1:L})$ in the discrete distribution for $s_t$; and 3) addition of a prior for $\{ \gamma_\ell \}$. We again favor simplicity and assign independent Bernoulli($\pi^\gamma_\ell$) priors to each $\gamma_\ell$. \color{black} One option is to set $\pi_\ell^\gamma$ equal to a constant for all lags, a common choice for variable selection in regression settings. When modeling nonlinear dynamics, however, subsets of lags are often highly correlated and subject to aliasing. We thus prefer to use, as a default, a decreasing sequence for $\pi^\gamma_\ell$ that helps identify the model by giving ordered preference to lower lags. As a specific choice, $\pi^\gamma_\ell= 0.1 + (0.4/0.5) * 0.5^\ell$, for $\ell=1, \ldots, L$, geometrically decreases from 0.5 to 0.1 to promote sparsity and dimension reduction. 
Supplement \ref{sec:appendix_sensitivity} explores posterior sensitivity to these prior options.
\color{black}

\subsection{Posterior inference}
\label{sec:inference_lagselect}

The proposed setup is minimally disruptive to the MCMC algorithm outlined in Section \ref{sec:computation}. Conditional on $\bm{\gamma}$, the effect of selection on the mixture kernels, and hence most of the Gibbs updates, is straightforward. We update $\bm{\gamma}$ as a block, with a Metropolis step that proposes switching a random subset of $\bm{\gamma}$ (similar to Section 3.3 of \citealp{schafer2013sequential}).
Details are given in Supplement \ref{sec:appendix_lagSelection_globalPost}. 
\color{black}
Although the $\bm{\gamma}$ update has computational complexity on the same order as that of $\{ \bm{\eta}^x_h \}$, the proposed method saves elsewhere by reducing the effective number of lags ($L$) in other updates.
\color{black}

It is well known that variable selection methods of this type tend to result in slowly mixing MCMC algorithms \citep{ohara2009review}. Proposed changes in $\bm{\gamma}$ are often incongruous with current-state values of model parameters, which are shared across selection configurations. Furthermore, when $\gamma_\ell = 0$, draws for the associated parameters revert to their prior distributions, which may be diffuse relative to their posterior distributions when $\gamma_\ell = 1$, producing draws that will discourage returning to $\gamma_\ell = 1$. 
Alternative methods such as Gibbs variable selection \citep{dellaportas2002gvs} adapt the prior to improve mixing, but require tuning. We do not pursue this here, but note that despite mixing difficulties and attenuated posterior probabilities for alternate lag configurations, our experience has been that MCMC chains can provide useful inferences. 
We recommend running multiple MCMC chains, initialized at different selection configurations. We begin MCMC with a phase in which $\bm{\gamma}$ is not updated, followed by iterated tuning or adaptation and burn-in phases with the full sampler, followed by a final burn-in.

Posterior inferences for relevant lags from MCMC samples are trivial, requiring only samples of $\bm{\gamma}$, which can be aggregated across iterations to obtain a posterior probability of inclusion for each lag. 
%
%
The full expression for the transition density, marginalizing over all $2^L$ possible lag configurations, is
\begin{align}
  \label{eq:condmodel_trunc_marggamma}
  \tilde{f}_{Y|X}(y\mid\bm{x}) = \sum_{\bm{\gamma} \in \{0,1\}^L} \sum_{h=1}^H \tilde{q}_h(\bm{x} \mid \bm{\gamma}) \, \Ndist_{(h)}(y\mid \mu(\bm{x} \mid \bm{\gamma}), \sigma^2 ) \, \Pr(\bm{\gamma}) \, ,
\end{align}
where $\Pr(\bm{\gamma})$ can refer to either the prior or marginal posterior of $\bm{\gamma}$. 
In practice, we bypass the burdensome outer summation in (\ref{eq:condmodel_trunc_marggamma}) and instead calculate the lag-conditional version of the transition density in Section \ref{sec:transdens_estimation} across MCMC samples, which yields the desired posterior inferences marginalized with respect to the posterior of all model parameters.

Conditional on lag selection, posterior inference for functionals proceeds as in Section \ref{sec:transdens_estimation}, with appropriate modifications to include $\bm{\gamma}$ 
(see Supplement \ref{sec:appendix_lagSelection_globalFunctionals} for details). 
Calculation of transition density and mean estimates requires the full $\bm{x} \in \mathbb{R}^{L}$, regardless of inferences for $\bm{\gamma}$. However, one may be interested in inferences conditional on a certain lag configuration, or marginal inferences that in some way ignore or average over the effect of a subset of $\bm{x}$. Suppose one has fit a model with $L=3$ and desires to examine the transition mean function of the first two lags only when $\bm{\gamma} = (1,1,0)$. One option would be to use only posterior samples for which this lag configuration was active (taking into account the order of full-conditional sampling), 
given a sufficiently long MCMC chain. They may then calculate the $\bm{\gamma}$-modified transition density using these samples for any $\bm{x}$, substituting a dummy or default value in for $x_3$, and examining the transition density or mean as a function of $x_1$ and $x_2$ only. If fewer than all posterior samples coincide with a particular configuration, one may proceed in the same way, substituting default (or average) values in for elements of $\bm{x}$ hypothesized to be inactive and examining inferences (calculated from posterior samples, including $\bm{\gamma}$) as a function of the subset of interest. We caution that using a subset of samples ignores posterior uncertainty, and that one should test the resulting inferences for sensitivity to the default values used for inactive $x_\ell$ before making conclusions. For example, one could change the default values in $\bm{x}$, or replace them with random values drawn uniformly across the range of $\{ y_t \}$, as demonstrated in Section \ref{sec:illustrations}.

\subsection{Local lag selection}
\label{sec:local_lagsel}

Thus far, we have used a single set of global indicators,  $\{ \gamma_\ell \}$. 
If one believes that lag (variable) dependence varies across the predictor space $\mathbb{R}^L$, it is straightforward to instead use a separate set $ \{ \gamma_\ell^{(h)} \} $ for each mixture component 
$h = 1, \ldots, H$, in which case the indicators become part of $\bm{\eta}_h$. 
Model extensions and implementation for local lag selection require only slight modifications to the procedures in Sections \ref{sec:lagselection_extensions} and \ref{sec:inference_lagselect}. Mixture kernels $\Ndist_{(h)}(y_t \mid \bm{y}_{t-1:L})$ and weight kernels $\Ndist_{(h)}(\bm{y}_{t-1:L})$ are modified in the same manner as before, but use a unique $\bm{\gamma}_h = \left( \gamma_1^{(h)}, \ldots , \gamma_L^{(h)} \right)$ for each $h = 1, \ldots, H$. With replicates of each $\gamma_\ell$ across components, the $L$ independent Bernoulli priors become part of $G_0$, and we assign independent mixture priors for each $\pi^\gamma_\ell$. Following \citet{chung2009probitsb_ssvs} and \citet{lucas2006sparse}, we use $\pi^\gamma_\ell \sim ( 1 - \pi^\pi_{\ell} ) \, \delta_0( \pi^\gamma_\ell ) + \pi^\pi_{\ell} \, \Betadist( \pi^{\gamma}_\ell \mid a^{\pi}_\ell, b^{\pi}_\ell ) $, for $\ell = 1, \ldots, L$, where $\pi^\pi_{\ell} \in (0,1)$, and $\delta_0$ is the Dirac delta measure centered at 0. We use $\pi^\pi_{\ell} = 0.1 + (0.4/0.5) * 0.5^\ell$, for $\ell=1, \ldots, L$, $a^\pi_1 = \cdots = a^\pi_L = 1.0$, and $b^\pi_1 = \cdots = b^\pi_L = 0.5$ as default values.

All modifications to MCMC updates in Section \ref{sec:inference_lagselect} still apply, but require mixture weight and kernel calculations to reference their respective $\bm{\gamma}_h$; 
see Supplement \ref{sec:appendix_lagSelection_localPost} for details. 
\color{black}
Local lag selection modestly increases computational complexity as well as MCMC runtime relative to global selection (see Supplement \ref{sec:appendix_sensitivity}). This is due to $H$ repeated calculations of the weight denominator across all observations (each requiring up to $O(THL^3)$, or $O(THL)$ for diagonal $\bm{\Sigma}^x$, operations). In our experience, however, increased MCMC efficiency renders local selection worthwhile.
\color{black}

Inference for global dependence can be assessed with local selection, but is more nuanced. 
We assess global lag dependence by monitoring the weight $ \sum_h \gamma^{(h)}_\ell \, \sum_t 1_{(s_t = h)} / (T - L) $, which gives the proportion of observations in the time series belonging to mixture components for which lag $\ell$ is active. Alternatively, we can replace $\gamma_\ell^{(h)}$ in the preceding expression with $\gamma_\ell^{(h)} 1_{(\lvert \beta_\ell^{y\, (h)} \rvert > b_0)}$, 
for some small threshold $b_0 > 0$, requiring both dependence in the weights {\em and} a minimum contribution to the slope of the kernel for a lag to be considered active. The quantities $ \sum_h \omega_h \, \gamma^{(h)}_\ell$ and $\pi^{\gamma}_\ell$ are also informative. Inferences for transition densities and associated functionals again follow the procedures in Section \ref{sec:inference_lagselect}.

\subsection{Data illustrations incorporating lag selection}
\label{sec:illustrations_lagselect}

We now revisit the analyses from Section \ref{sec:illustrations} with lag selection, and include two additional examples. All models in this section utilize diagonal $\bm{\Sigma}^x$ and default prior settings. 
\textcolor{black}{Parameters of the components of $G_{0}$ associated with 
$\bm{\eta}^y$ were fixed.}
For each example, four MCMC chains were randomly initialized using the strategy described in Section \ref{sec:computation}. Two chains were initialized with all lags off and two were initialized with all lags on. Tuning stages were followed by 300,000 burn-in samples. The next 500,000 iterations were then thinned to 5,000 for inference (and further thinned for computationally expensive functionals such as surfaces). Both global and local lag selection were employed and compared. 

\subsubsection*{Simulated data: linear autoregression}

To test the model's ability to identify simple structure, for which the proposed model is over-specified, we generated time series from a stationary, 
\textcolor{black}{Gaussian}
linear autoregressive model of order two. 
Models with global and local lag selection perform well on time series of varying length, successfully recovering parameter values and decisively selecting the first two lags (with non-negligible inclusion of lag 3 for the longer series). 
Further details are given in 
Supplement \ref{sec:appendix_lagSelection_AR2}.

\subsubsection{Simulated data: Ricker model}

Model runs ($T=75$, $L=5$, $H=40$) fit to the nonlinear simulation from Section \ref{sec:ex_appdataLag2} consistently recover lag dependence as well as the nonlinear dynamics with both global and local selection. Specifically, lag 2 is consistently kept on for all occupied mixture components throughout the chains, and other lags are generally off, with greater mixing in the model with local selection. Inferences appear fairly robust to choice of prior signal-to-noise ratio $\mathcal{R} \in \{ 5.0, 8.0 \}$. The estimated transition mean functional, on lag 2 only, is visually very similar to the left panel of Figure \ref{fig:appdata_lag2only_transMean}. A few runs include lag 4, which is reasonable given that the data reside in two diagonal quadrants of the $(y_{t-2}, y_{t-4})$ lag embedding space.

    

\subsubsection{Old Faithful data}

Model runs ($T=294$, $L=5$, $H=40$) fit to the Old Faithful time series have mixed results. Global lag selection runs with low prior 
\textcolor{black}{signal-to-noise ratio}
$\mathcal{R} = 5.0$ all converge to lag 1 only with no mixing over other configurations. Runs with $\mathcal{R} = 8.0$ continue exploring selection configurations past the specified burn-in phase, on very long timescales. One run retains lag 3 and another uses lag 2 for part of the chain. Local selection yields similar results to global, but with exploration of lag inclusion on shorter time scales and some local inclusion of lags 2 and 3. Nevertheless, these runs do not detect the density dependence noted in Section \ref{sec:ex_oldFaithful}. Additional runs with priors more favorable to higher lags and larger prior weight-kernel variance also miss dependence on lag 2.


\subsubsection{Pink salmon data}

\begin{figure}[b!]
	\centering
	
	\begin{tabular}{cccc} 
		\multicolumn{4}{c}{\includegraphics[scale=0.55]{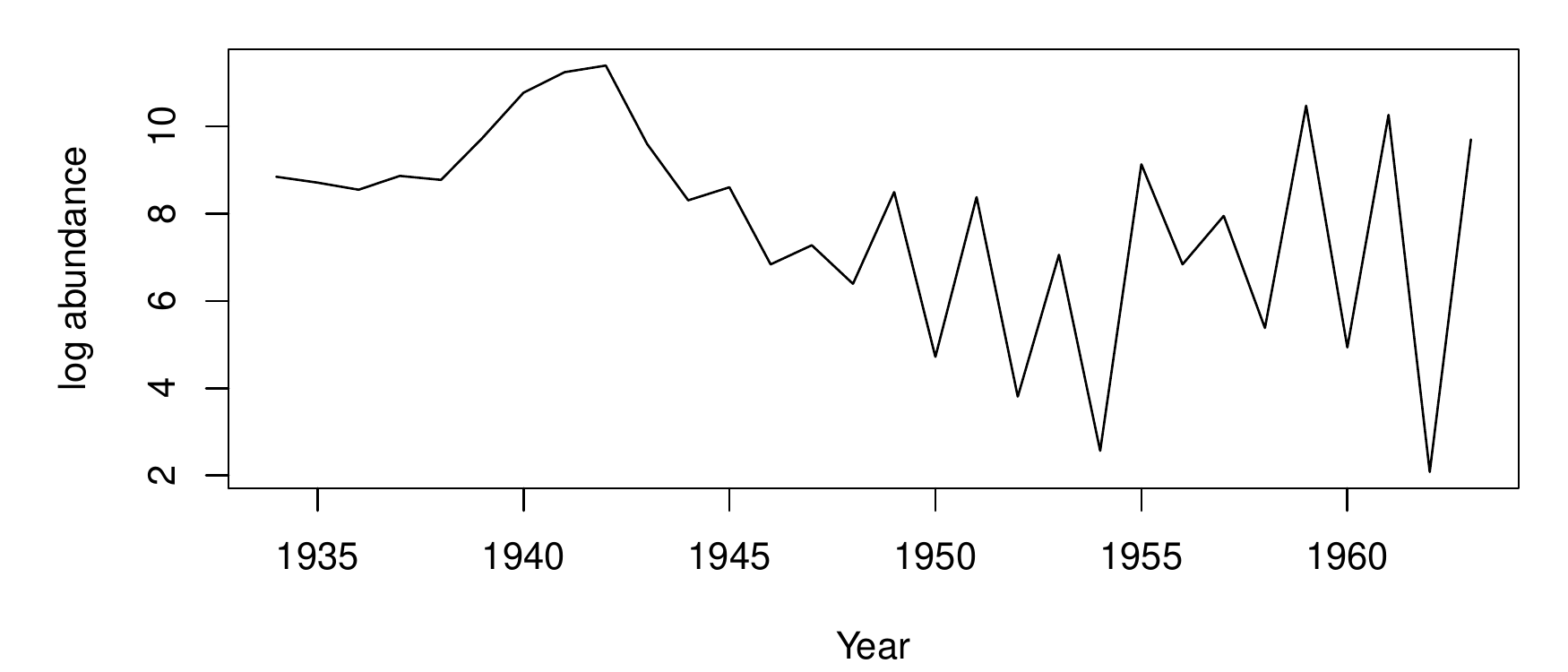}}
	\end{tabular}
	\caption{{\small Trace 
of the natural logarithm of pink salmon abundance in Alaska from 1934 to 1963.}} 
	\label{fig:pinksalmon_timeseries}
\end{figure}

We next investigate a time series of annual pink salmon abundance (escapement) in Alaska, U.S.A.\ 
 \citep{PinkSalmonData}, \color{black} whose life cycle 
reliably follows a two-year pattern \citep{heard1991pinksalmon}. Naive modeling of annual population dynamics based on the previous year only would capture inter-population, rather than generational dynamic dependence. 
\color{black}
We expect even lags to have the most influence in predicting the current year's population. The trace of the natural logarithm of abundance in Figure \ref{fig:pinksalmon_timeseries} suggests a comprehensive analysis might appropriately include non-stationarity with long-term trends, which we forego in favor of a simple demonstration. 
Lag scatter plots (not shown) suggest that we should be able to detect lag dependence structure, even with as few as 30 observations.

\begin{figure}[b!]
  \includegraphics[height=1.5in, trim = 5 0 15 0, clip]{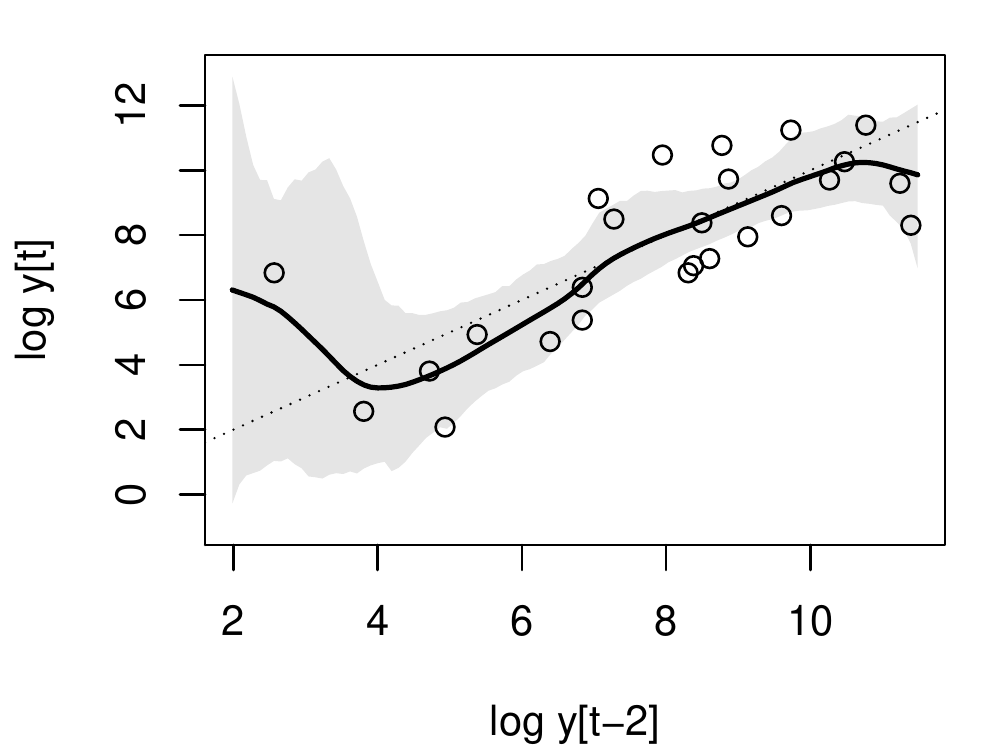}
  \includegraphics[height=1.5in, trim = 45 0 15 0, clip]{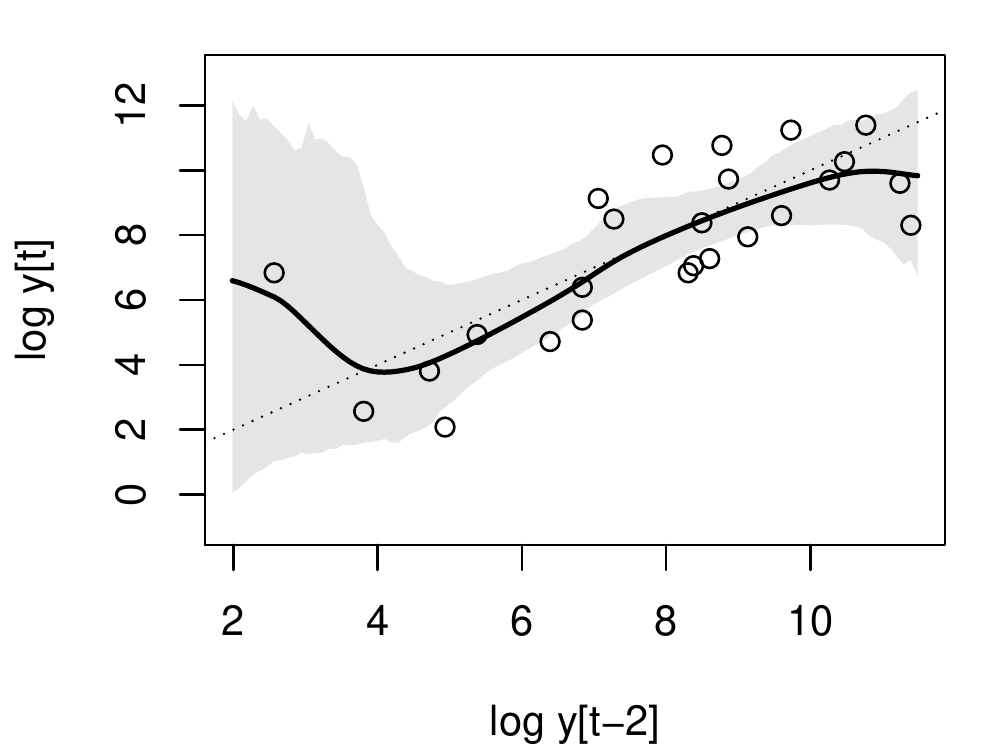}
  \includegraphics[height=1.5in, trim = 45 0 15 0, clip]{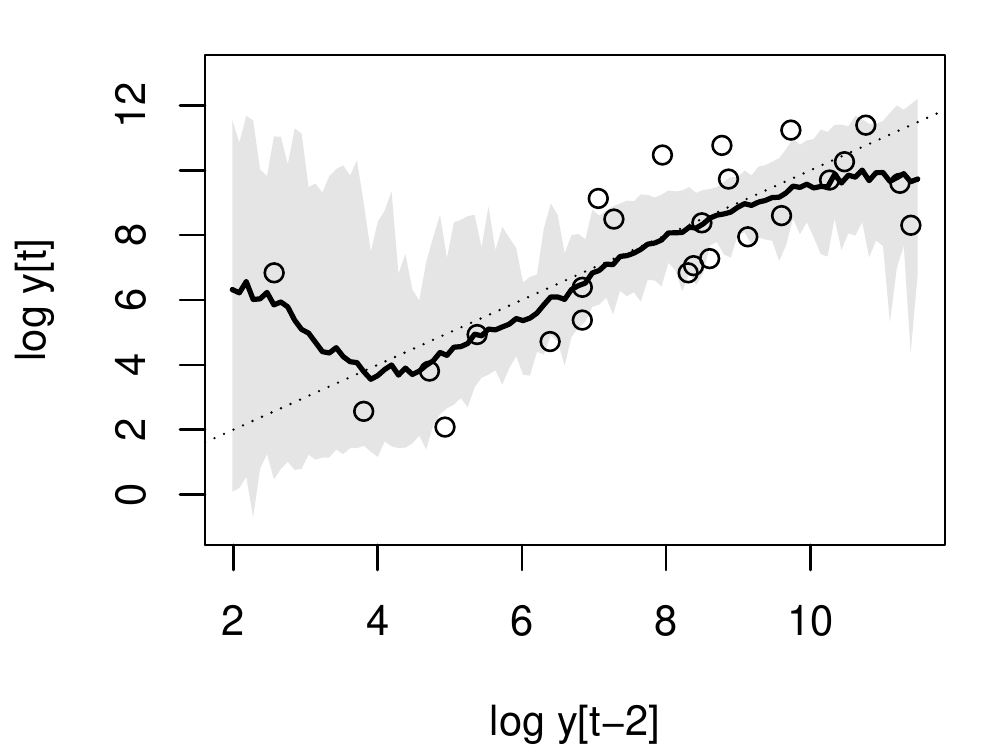}
\caption{{\small Model fit to the logarithm of annual pink salmon abundance with $T=30$, $L=5$, using global (left) and local (center and right) lag selection. In the global case, lag 2 is predominately selected, and this plot reflects only iterations of MCMC for which lag 2 is on. In the local case, other lags were fixed at the mean (center) or randomly drawn (right). The plots include pointwise posterior mean estimates and 95\% credible intervals for the transition mean as a function of lag 2, together with observed two-step transitions. The dotted reference line has unit slope and passes through the origin.}}
\label{fig:pink_lagsel_transMean}
\end{figure}

Model runs ($T=30$, $L=5$, $H=25$) fit to the pink salmon data demonstrate sensitivity to prior and model specification. Most runs with global lag selection and higher prior signal-to-noise ratio ($\mathcal{R}=8.0$) deselect all lags, although one run has lag 2 active for many inference samples. Runs with lower $\mathcal{R} = 5.0$ deselect all lags except lag 2, which is on for long periods in three of four chains. Local lag selection consistently retains lag 2 throughout most inference samples, as well as lag 4 occasionally. Increasing $\mathcal{R}$ tends to result in a higher inclusion probability for lag 4, presumably from a tendency to over-fit a transition surface informed by data only in diagonal quadrants of the $(\log(y_{t-2}), \log(y_{t-4}))$ space, similar to the Ricker model above. Figure \ref{fig:pink_lagsel_transMean} reports posterior inferences for the transition mean as a function of lag 2, 
%
\textcolor{black}{under the global and local lag selection model versions.}
Inferences for the transition mean as a function of lag 2 only appears appropriate and insensitive to other lags.


\section{Transition density estimation performance}
\label{sec:forec_comparison}

Transition density estimation is a primary objective of the methodology. To compare density estimation across model configurations and data scenarios, we fit the model to simulated time series exhibiting various features and evaluate Monte Carlo estimates of the Kullback-Leibler (K-L) divergence between the estimated and true transition densities.

\begin{figure}[t!]
  \includegraphics[width=2.25in]{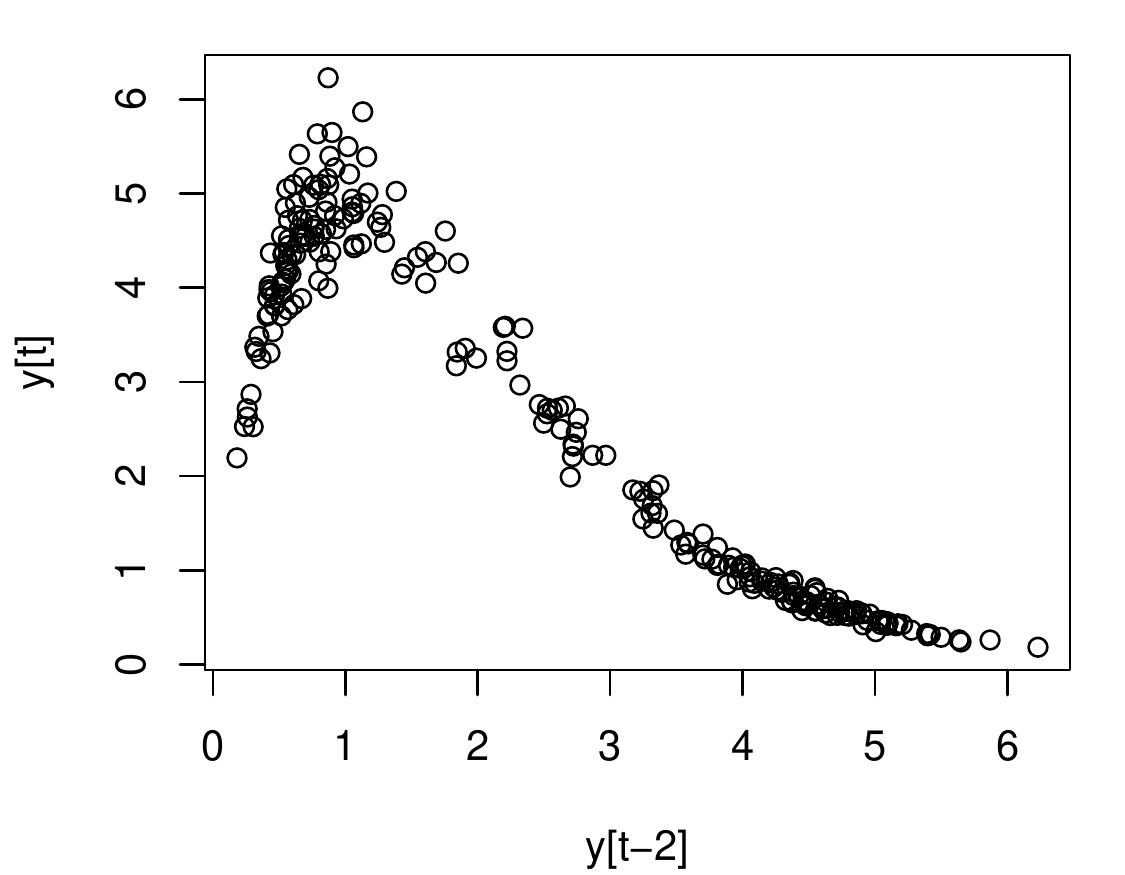}
\caption{{\small Lag scatter plot from the modified single-lag nonlinear simulation with log-normal transition density.}}
\label{fig:appdata2_hsced}
\end{figure}

\begin{figure}[t!]
  \includegraphics[width=2.4in]{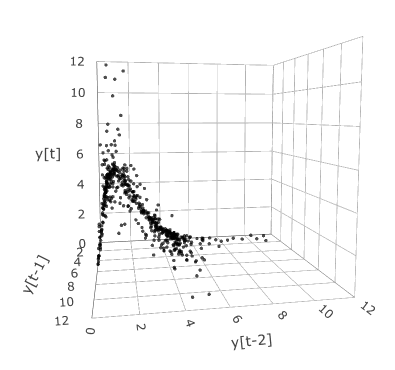} \includegraphics[width=2.75in]{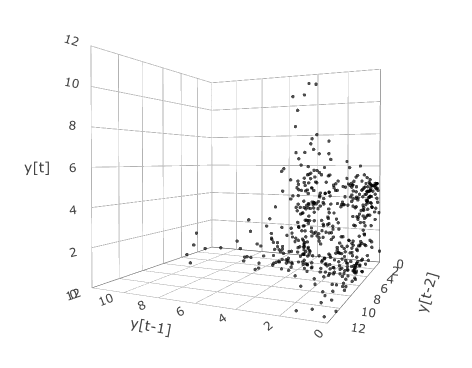}
\caption{{\small Lag scatter plot from the modified two-lag nonlinear simulation with log-normal transition density.}}
\label{fig:appdata2_condhsced}
\end{figure}

The simulated time series are variants of the Ricker-type system in (\ref{eq:appDatalag2gen}). The first modification replaces the additive Gaussian error with multiplicative log-normal error. Specifically, transitions were generated from
\begin{align}
  \label{eq:appdata2_hsced}
  y_t = y_{t-2} \exp(2.6 - y_{t-2} + \epsilon_t) \, , \qquad \epsilon_t \simiid \Ndist(0, (0.09)^2) \, ,
\end{align}
corresponding to a log-normal transition density. 
This produces right skew and heteroscedasticity in the transition distribution, which continues to depend exclusively on the second lag. The lag scatter plot in Figure \ref{fig:appdata2_hsced} depicts 250 transitions. We refer to this modification as the single-lag, log-normal simulation. The second modification adds dependence on the first lag through the log-scale, which is equal to $0.09 \, y_{t-1}$. Thus the transition distribution is still log-normal, with each parameter depending on a separate lag. The lag scatter plot in Figure \ref{fig:appdata2_condhsced} depicts 500 transitions, demonstrating dependence of the variance on both lags. We refer to this modification as the two-lag, log-normal simulation. In all simulations, a sequence of 1,000 
observations was reserved for model fitting, and a validation set of size 1,000 was randomly sampled from the subsequent 9,000 observations. In similar data scenarios with right skew and positive-valued variables, we have previously modeled observations on the logarithmic scale. We nevertheless proceed by fitting these series directly in order to study and compare how the proposed models handle heteroscedasticity, subtle departures from Gaussianity, and subtle variation in lag dependence.

The following models were fit using default settings to all three series: the proposed model (which we denote as the BNP-WMAR model, for Bayesian nonparametric, weighted mixture of autoregressive models) with $L=2$, full $\bm{\Sigma}^x$, and no lag selection; the BNP-WMAR model with $L=5$, diagonal $\bm{\Sigma}^x$, and global lag selection; and finally with $L=5$, diagonal $\bm{\Sigma}^x$, and local lag selection. Three chains were run for 
\textcolor{black}{the base model that does not incorporate lag selection,}
and four chains were run for each model with lag selection, with two chains initialized with all lags off and the other two initialized with all lags on. 

Each posterior sample was used to create density ordinates, denoted $\hat{p}(y_t \mid \bm{y}_{t-1:L})$ and calculated from the transition density in Section \ref{sec:transdens_estimation}, appropriately modified by lag selection indicators. With 2,000 replicate simulation draws $\{ y_j^{(i)} \}_{i=1}^{2{,}000}$ 
from the data-generating distribution (with density $p_{\text{true}}(y_t \mid \bm{y}_{t-1:L})$) for each validation pair $\{ (y_{j}, \bm{x}_j ) \}_{j=1}^{1{,}000}$, we approximated the Kullback-Leibler divergence using
\begin{align}
  \label{eq:KL}
  \begin{split}
   \KL(p_{\text{true}} \parallel \hat{p}) &\equiv \int p_{\text{true}}(y \mid \bm{x}) \log \left( \frac{p_{\text{true}}(y \mid \bm{x})}{\hat{p}(y \mid \bm{x})} \right) \diff y  \\
   &\approx \sum_{i=1}^{2,000} \log\left( p_{\text{true}}( y^{(i)} \mid \bm{x})  \right) - \log \left(  \hat{p}( y^{(i)} \mid \bm{x})  \right) \, ,
    \end{split}
\end{align}
averaged over validation observations and posterior simulations. Let $\hat{\mathrm{D}}_{\mathrm{KL}}$ denote the result. This loss metric is reported in Table \ref{tab:KL_compare} for two chains of each model fit to time-series of lengths $T=75$ and $T=305$ ($T=72$ and 302 for the model with $L=2$; using the same 70 and 300 observations used to fit the models with $L=5$). The two reported runs are those producing the minimum and maximum observed K-L divergence within each set.

In the single-lag, normal scenario, the burden of fitting an unnecessary dimension of the phase space is evident, particularly with the short time series. In the $T=302$ case with $L=2$, two of the chains use three mixture components, whereas one uses four and performs comparably to the models with lag selection. Both global and local lag selection perform well for both sample sizes, and yield accurate inferences for lag dependence, with occasional inclusion of lag 4.

\begin{table}[t]
	\centering
	\caption{{\small Comparison of 
	transition density estimation performance, measured by K-L divergence, from model fits to three simulations and with two sample sizes. The numbers in parentheses are $L$, the number of lags considered in each fit. Within each set, the minimum (left) and maximum (right) losses across runs are reported.}}
	\label{tab:KL_compare}
	\small

  \begin{tabular}{l l r r c r r}
    \toprule
    & & \multicolumn{5}{c}{ $\hat{\mathrm{D}}_{\mathrm{KL}}$ } \\
    \cmidrule{3-7}
        Simulation & Model & \multicolumn{2}{c}{70 obs.} & & \multicolumn{2}{c}{300 obs.}\\
    \midrule
    Single-lag, & Base model (2) & 2.756 & 3.761 &   & 0.273 & 0.384 \\
     \ \ normal & Global selection (5) & {\bf 0.777} & 0.821 &   & {\bf 0.239} & 0.252 \\
      & Local selection (5) & 0.792 & 0.828 &   & 0.250 & 0.264 \\
      & & & & & & \\
    Single-lag, & Base model (2) & 1.110 & 1.240 &    & 0.337 &  0.340 \\
     \ \ log-normal & Global selection (5) & {\bf 0.700} & 0.733 &    & { 0.296} & 0.326 \\
      & Local selection (5) & 0.723 & 0.776 &    & {\bf 0.296} & 0.305 \\
      & & & & & & \\
    Two-lag, & Base model (2) & 2.210 & 2.672 &    & 1.084 & 1.103 \\
     \ \ log-normal & Global selection (5) & 1.429 & 2.096 &    & 0.966 & 2.002 \\
      & Local selection (5) & {\bf 1.417} & 1.445 &    & {\bf 0.948} & 0.978  \\
        \bottomrule
    \end{tabular}  
  
\end{table}

Fitting two lags again hinders the base model in the single-lag, log-normal scenario when sample size is small. Both global and local lag selection perform well, with the best run of local selection only slightly outperforming the best run of global selection in the $T=305$ case. In the large sample, chains of the global selection model initialized with all lags on retain both lags 2 and 4, whereas chains initialized with all lags off retain only lag 2. Despite this lack of mixing, the discrepancy in K-L loss is minimal.

In the two-lag, log-normal scenario, the base model with $L=2$ has the advantage of being fixed at the correct lag structure. However, both models with lag selection manage superior performance. In the small sample, the base model over-fits a few points in the sparse region with $y_{t-1}$ low and $y_{t-2}$ high, producing inferences that fail to generalize. In contrast, the global selection model retains only lag 2 in three of four runs (selecting none in the poorly performing run), avoiding the over-fitting issue at the expense of missing density dependence on the first lag. The model compensates with a right-skewed transition density when $y_{t-2}$ is low, irrespective of $y_{t-1}$. Local lag selection has similar behavior in the small sample. All models struggle in the region with small values of lag 1 and large values of lag 2, where the true density is far more concentrated than estimated. 

In the large sample, global selection is inconsistent, correctly retaining both of the first two lags when initialized with all lags on, but retaining no lags and lag 2 only in respective runs initialized with no lags on. The two runs with correct lag selection yield effective density estimation, capturing variance and skew dependence on $y_{t-1}$. Local lag selection does the same, with consistent performance across initializations. 

The dimension reduction and parsimony afforded by lag selection provide significant gains in density estimation, as measured by K-L distance, and make a strong case for the proposed model extensions. Global selection can be effective when the dependence structure is simple, but we generally recommend 
\textcolor{black}{the more versatile local selection model.}

\section{Discussion}
\label{sec:discussion}

We have developed a modeling framework for fully nonparametric, nonlinear autoregressive models targeted at estimating transition densities. The model extends existing single-lag counterparts and further offers inference for lag dependence. We have demonstrated the model's utility with simulated, geological, and ecological data examples with diverse objectives. The model allows users to relax restrictive characteristics of standard models, or softly specify such through prior settings, within a single model. 
%

Results from the base model are promising, faithfully capturing known or anticipated features in the data examples. 
Of course, current computing bottlenecks limit what can practically be accomplished. For example, complex dynamics call for many mixture components and high truncation level $H$, and computations for updating component-specific parameters are not readily distributable in our approach due to their appearance in the normalized weights. 

The modeling objectives of estimating flexible transition densities, accommodating nonlinear dynamics, and selecting active lags offer many degrees of freedom that in most cases will not be entirely identified with data alone. Decisions must be made, and correspondent behaviors encouraged through the prior settings. As such, we recommend completing a thorough 
exploratory analysis of data. 
We further advise that practitioners fit models with a variety of prior signal-to-noise ratio and flexibility (through $\alpha$ and possibly $\delta^x$) settings, each with multiple MCMC chains.

Several considerations can help guide which settings are appropriate for a given scenario. One that bears on lag selection is the interplay between noise and signal. A model attempting to fit noise may erroneously reach into higher dimensions. However, in the absence of noise, finding a high-dimensional structure is an objective of techniques such as time-delay embedding. 
Another consideration arises from correlation among lags, which can result in multiple distinct lag configurations that each produce comparably effective forecasts. This partially motivates our recommendation of decaying inclusion probabilities in the prior.

Inference for relevant lags remains practically challenging.
Our experience has been that results from the models with lag selection tend to exhibit prior sensitivity, a natural consequence of the flexibility discussed. 
We have also noted that in models with lag selection, mixing challenges intensify with increased time series length, which tends to sharpen posterior modes. This often manifests through kernel coefficients being estimated at small, nonzero magnitudes while corresponding lag selection indicators remain on. Local selection helps alleviate this issue by breaking up the samples informing multiple lag indicators, naturally tempering the posterior distributions and encouraging greater mixing. The cost of added versatility and improved mixing is a more intricate picture of lag importance in posterior analysis.

Although binary lag inclusion parameters are easy to interpret, they offer limited insight to relative contributions from active lags. Such contributions can be quantified for the mean transition function through functional decomposition, 
but this is less straightforward for transition densities. Ideally, weak dependence would manifest in the posterior probability of inclusion. Alternatively, we envision a framework that quantifies lag importance with shrinkage of continuous parameters. Continuous quantification of lag importance could in turn reduce the influence of weight kernels relative to the DP weights and thus accommodate multiple sources of influence on the mixture weights.

Notwithstanding theoretical and practical challenges, lag selection is critical for dimension reduction and is an integral part of this work. Simpler models can partially avoid some of the challenges noted, but risk failing to model, or even detect, nonlinear and/or non-Gaussian dynamics. Our proposed methods extend Bayesian nonparametric density autoregressive modeling by accommodating multiple lags and providing a framework for lag selection that works in concert with the other objectives.

\newpage

\bibliographystyle{jasa3.bst}
\bibliography{Bibliography}

\beginsupplement


\section{Model for stationary time series}
\label{sec:appendix_stationary}

One way to ensure stationarity of a process with transition kernel (\ref{eq:condmodel}) is to espouse the joint density interpretation of the mixture (\ref{eq:jointmodel_stickbreak}) 
and constrain $\bm{\mu}_h = \bm{1} \mu_h$ and $ \bm{\Sigma}_h $ to be positive definite Toeplitz, thereby ensuring reversibility with identical marginal distributions ($ \int_\mathcal{A} f_{Y,X}(y, \bm{x}) \diff y = \int_\mathcal{A} f_{Y,X}(y, \bm{x}) \diff x_L $ for any measurable set $\mathcal{A}$). This is the approach taken by \citet{antoniano2016} and \citet{kalli2018npbvar}, who assume an AR(1) covariance structure for $ \bm{\Sigma}_h $. 
Generally, each mixture component accommodates up to $L+2$ parameters. The Markovian likelihood then arising from (\ref{eq:condmodel}) precludes closed-form Gibbs sampling.

Model-based lag selection presents the primary challenge in a stationary model with $L>1$. An additional requirement is that lag selection is global. To see this, consider a two-component mixture of bivariate Gaussian densities, with each component using a separate lag: $f(y_t, y_{t-1}, y_{t-2}) = \omega \Ndist_{(1)}(y_{t}, y_{t-1}) + (1 - \omega) \Ndist_{(2)}(y_t, y_{t-2})$. Integrating out $y_t$ leaves a mixture of univariate Gaussian densities, whereas integrating out $y_{t-2}$ leaves one bivariate and one univariate component, yielding two distinct marginal densities for $(y_t, y_{t-1})$ and $(y_{t-1}, y_{t-2})$. We can maintain stationarity with a joint density defined over a set of nonconsecutive lags (e.g., $f(y_t, y_{t-2})$) if we assume that conditionally independent (in the transition) subsequences of the time series follow the same stationary distribution.

Given the structural requirement on $\bm{\Sigma}_h$, the Cholesky factorization in Section \ref{sec:CholFact} is not useful in the stationary case. We instead consider constructing $\bm{\Sigma}_h$ from autoregressive coefficients, which we denote $\bm{\phi}_h = ( \phi_1^{(h)} , \ldots, \phi_L^{(h)} ) $, allowing some control over lag dependence. Given $\bm{\phi}_h$, $\mu_h$, and innovation variance $\sigma^2_h$, we can recover $\bm{\Sigma}_h$ by solving the set of homogeneous difference equations defining the stationary autocovariance (equivalently, the Yule-Walker equations). To maintain causality (a sufficient condition for stationarity) of each mixture component, all roots of the AR characteristic polynomial must lie outside the unit circle \citep[pp.\ 86, 95, 113]{shumwaystoffer2017}.

We briefly note four possible approaches to modeling lag dependence with $\bm{\phi}_h$:

\begin{enumerate}
\item
Specify a full $\bm{\phi}_h$ vector for each mixture component and use $ \{ \gamma_\ell \, \phi^{(h)}_\ell \}_{\ell = 1}^L $ to construct $\bm{\Sigma}_h$, as in Section \ref{sec:lagselection}. A drawback of this approach is that the weight kernels remain $L$-variate Gaussian densities, dependent on all $L$ lags.

\item
To eliminate dependence on inactive lags, use the original $\bm{\phi}_h$ to construct $\bm{\Sigma}_h$ before marginalizing both mixture and weight kernels as $\int \Ndist_{(h)}(y, \bm{y}_{t-1:L}) \, \diff \bar{\bm{y}}_{t-1}$ and $\int \Ndist_{(h)}(\bm{y}_{t-1:L}) \, \diff \bar{\bm{y}}_{t-1}$, respectively, where $\bar{\bm{y}}_{t-1}$ contains the $y_{t - \ell}$ for which $\gamma_\ell = 0$. A drawback of this approach is that the model is over-parameterized, leaving unidentified parameters and possibly inflating uncertainty.

\item
Combine approaches 1 and 2, removing the effect of inactive lags when constructing the transition density, with joint densities defined over possibly nonconsecutive lags. This approach seems the most promising.

\item
One could define the component-specific parameter space as a union of sets of distinct AR coefficient vectors, and perform transdimensional MCMC. Beyond the computational burden this would create with $H$ mixture components, the combinatorial complexity of considering subsets of $\{ \phi^{(h)}_\ell \}_{\ell = 1}^L$ makes this option unattractive.

\end{enumerate}

\noindent In all cases, it would be necessary to check the roots of characteristic polynomials prior to accepting any candidate $\bm{\phi}$ or $\bm{\gamma}$ during MCMC, and mixing could suffer from inability to marginalize over mixture kernel parameters. One way to avoid checking roots is to work directly on the space of characteristic polynomial roots, as in \citet{huerta1999priors}, although they disallow gaps in lag dependence and utilize reversible-jump algorithms to identify model order. Nevertheless, if stationarity is required in a particular modeling scenario, it may prove worthwhile to consider these approaches.

\section{Prior settings}
\label{sec:appendix_priorsettings}

\begin{figure}[t!]
    \begin{center}
      \begin{tabular} {ll}
        \includegraphics[width=2.5in]{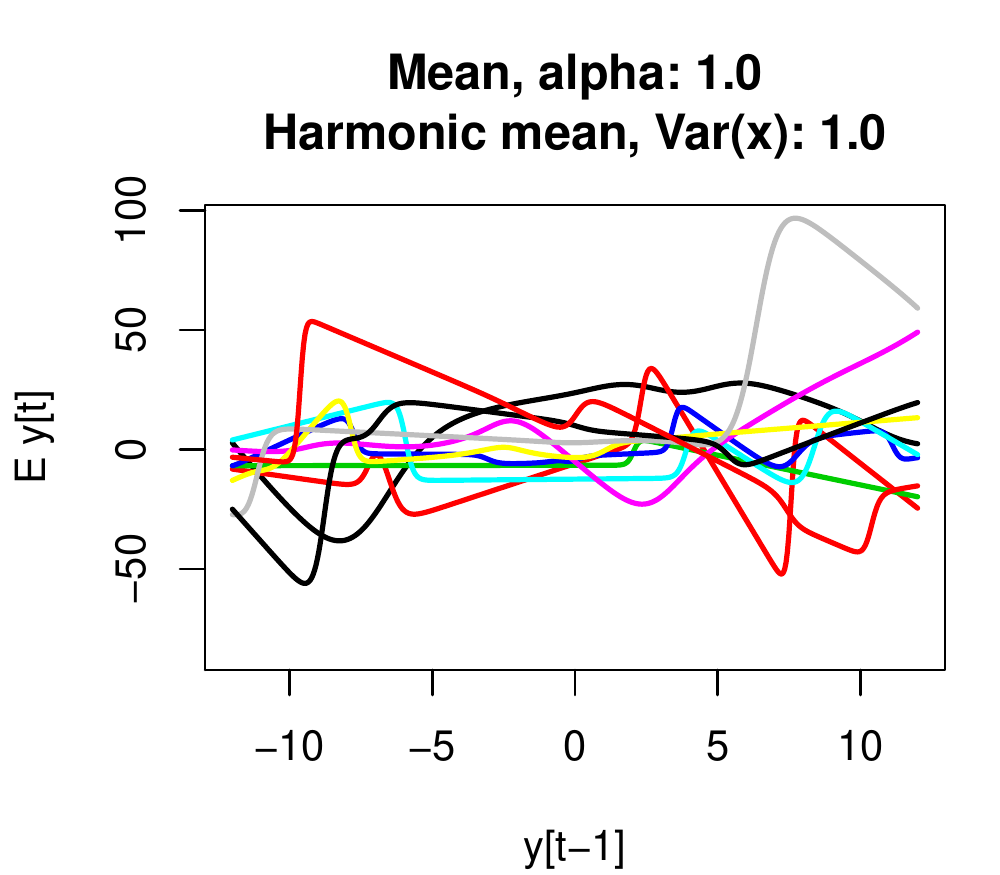} & 
        \includegraphics[width=2.5in]{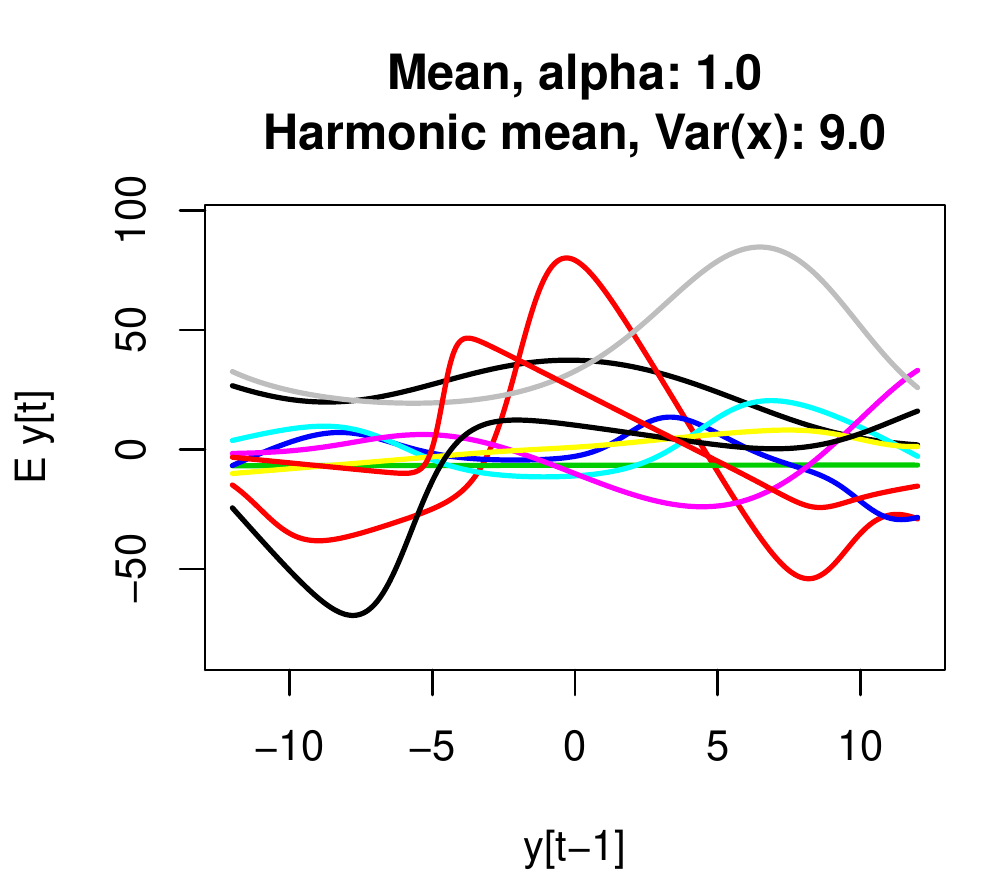} \\ 
        \includegraphics[width=2.5in]{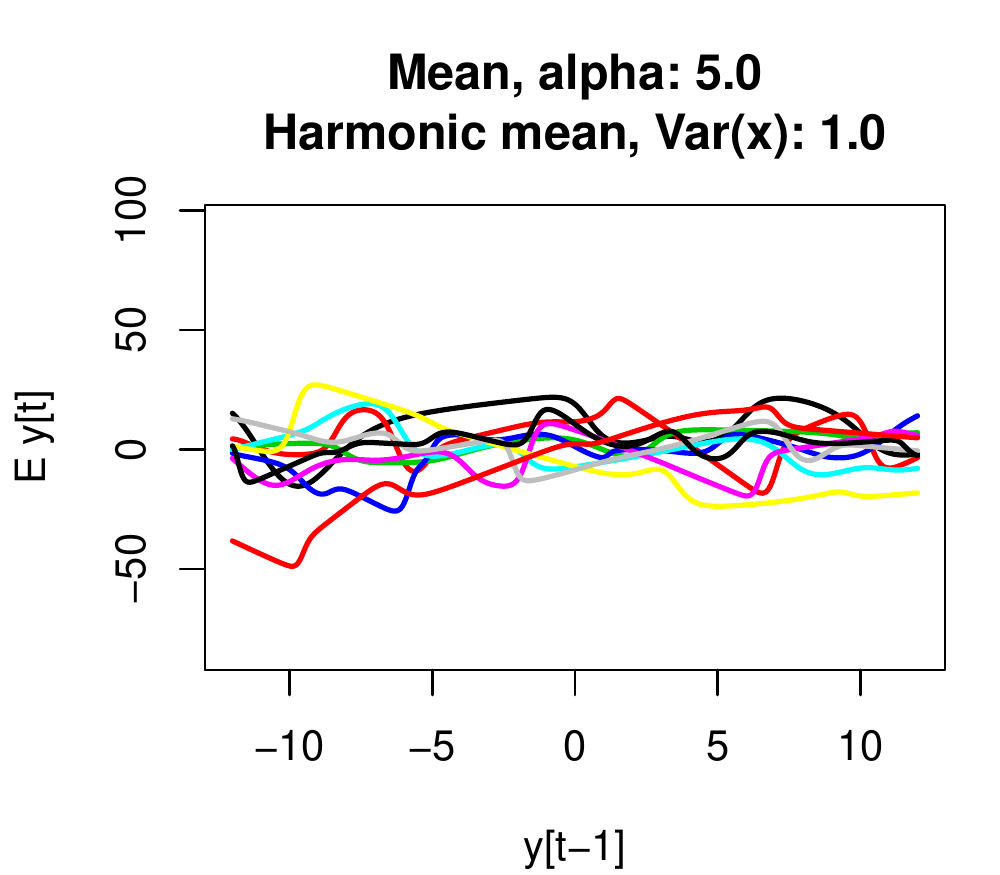} &
        \includegraphics[width=2.5in]{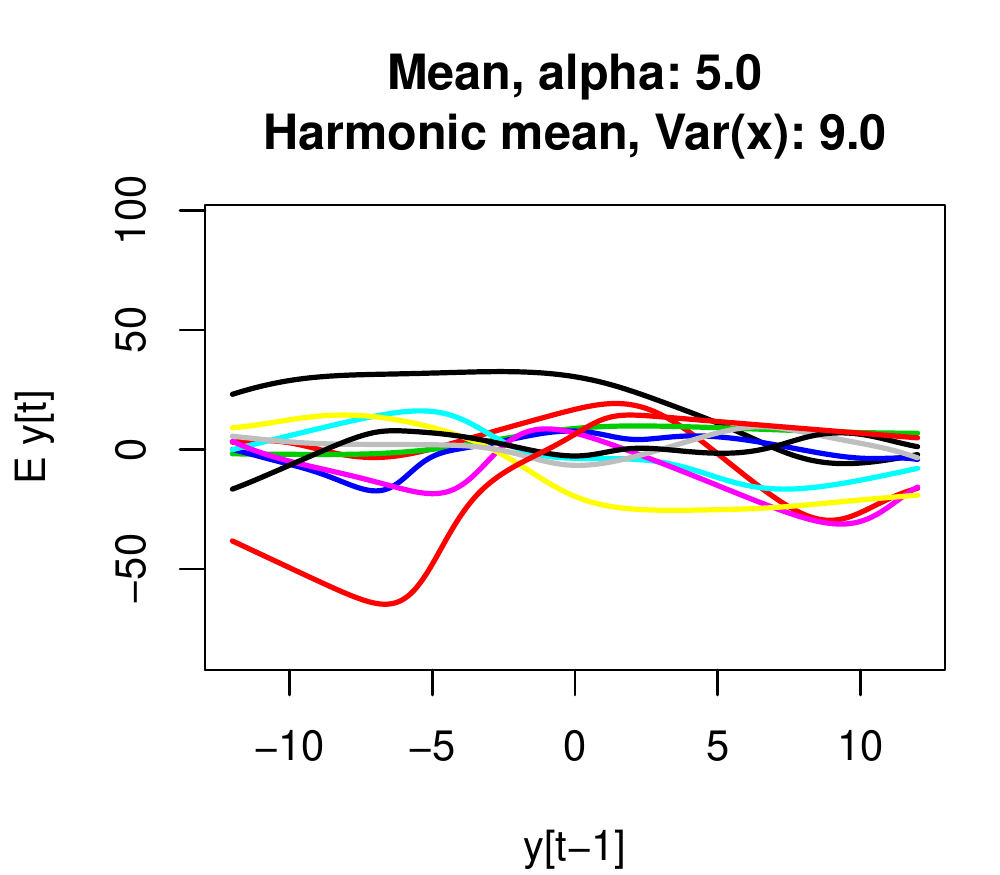}
      \end{tabular}
      
    \end{center}
    \caption{Ten prior realizations of the transition mean for the proposed nonparametric model with a single lag, under combinations of prior settings for $\alpha$ and $\delta^x$.
    \label{fig:prior_Ey}}
  \end{figure}

  We can visualize the effects of prior settings through prior simulation in low-dimensional models. As an example, Figure \ref{fig:prior_Ey} depicts several realizations of the transition mean for a model with a single lag. The realizations are drawn under combinations of prior settings for $\alpha$ (through the shape parameter with the scale fixed at $1.0$) and $\delta^x$ (through the prior mean of $s_{0}^x$). Restricting the number of components with low values of $\alpha$ results in transition mean functions with few change points and long stretches of near linearity, whereas allowing more components increases variability in the curve. Low values for $\delta^x$ likewise encourage rigid transition mean curves with abrupt change points. Increasing the variance in the weight kernels has a smoothing effect, as expected. 
  Note that in regions of the lag space where multiple mixture components carry significant weight, the transition density can be multimodal, with a transition mean that does not closely follow any one of the component-specific lines.
  As with the transition mean, one can use prior simulation to elucidate the effects of prior settings for transition densities to aid practitioners in specifying desired characteristics and performing sensitivity analyses.

\section{\textcolor{black}{Computing time and sensitivity analysis}}
\label{sec:appendix_sensitivity}

We have noted sensitivity of results to model and prior settings at multiple instances. While model flexibility can be a feature, we underscore the importance of understanding the effects of certain settings on model performance and posterior inferences. This section reports a sensitivity analysis using simulated data from the Ricker model (Sections \ref{sec:ex_appdataLag2} and \ref{sec:illustrations_lagselect}) based on a $2^6$ factorial experiment with the factors listed in Table \ref{tab:sensitivity_factors}. Each treatment combination includes two replicates, yielding 128 model runs in total. We measured the following diagnostic responses: MCMC run timing, MCMC convergence issues, posterior summaries of lag inclusion, number of occupied mixture components, interval-estimate width for the transition mean functional, and mean-squared error. These results do not necessarily generalize to other data sets, whose complexity and order of dependence can affect the monitored responses. However, they do confirm intuition on the roles of these factors.

\begin{table}[b!]
	\centering
	\caption{Six factors used in the sensitivity analysis, including values for the two levels used for each. $\Gammadist(a,b)$ denotes a gamma distribution with mean $a/b$. Sequences of prior lag-inclusion probabilities are either geometrically decreasing from 0.5 (first lag) to 0.1 (lag $L$), or constant at 0.5.}
	\label{tab:sensitivity_factors}

	\begin{tabular}{l r r}
	\toprule
		
		Factor & \multicolumn{2}{c}{Levels} \\

	\midrule

		Sample size, $n$ & 50 & 150 \\
		Lag-selection method & global & local \\
		Prior signal-to-noise ratio, $\mathcal{R}$ & 4 & 12 \\
		Prior for DP concentration, $\alpha$ & $\Gammadist(10, 10)$ & $\Gammadist(60, 8)$ \\
		Prior lag-inclusion prob.\ $\{ \pi^\gamma_\ell \}$ (or $\{ \pi^\pi_\ell \}$) & 0.5 $\searrow$ 0.1 & 0.5 \\
		Initial value for $\bm{\gamma}$ (or $\bm{\gamma}_h$) & $\bm{0}$ (none) & $\bm{1}$ (all) \\
	
	\bottomrule
	\end{tabular}

\end{table}

Aside from the controlled factors, default settings were used for the models, including $L=5$ and diagonal $\bm{\Sigma}^x$. The DP truncation was set at an overly conservative $H=80$, which appears to be sufficiently high to accommodate the prior for high $\alpha$. Each chain was run for approximately 1,150,000 iterations, with the first 400,000 discarded as burn-in. Posterior means for $\alpha$ concentrate below 0.5 under the $\Gammadist(10, 10)$ prior and around 4.5-6.5 under the $\Gammadist(60, 8)$ prior, indicating that, for small to moderate sample sizes, this prior choice is essentially a model setting with only mild influence from data.

\subsection*{Timing}

Timing and sensitivity runs were performed on single-core, 64-bit Intel  \textregistered \ Xeon \textregistered \ Gold 6248 processors running at 2.50 GHz, using our package in \textit{Julia} Version 1.4.1 \citep{bezanson2017julia}. We report on the period after adaptation and burn-in.

\begin{figure}[tb]
    \begin{center}
        \includegraphics[width=6.5in]{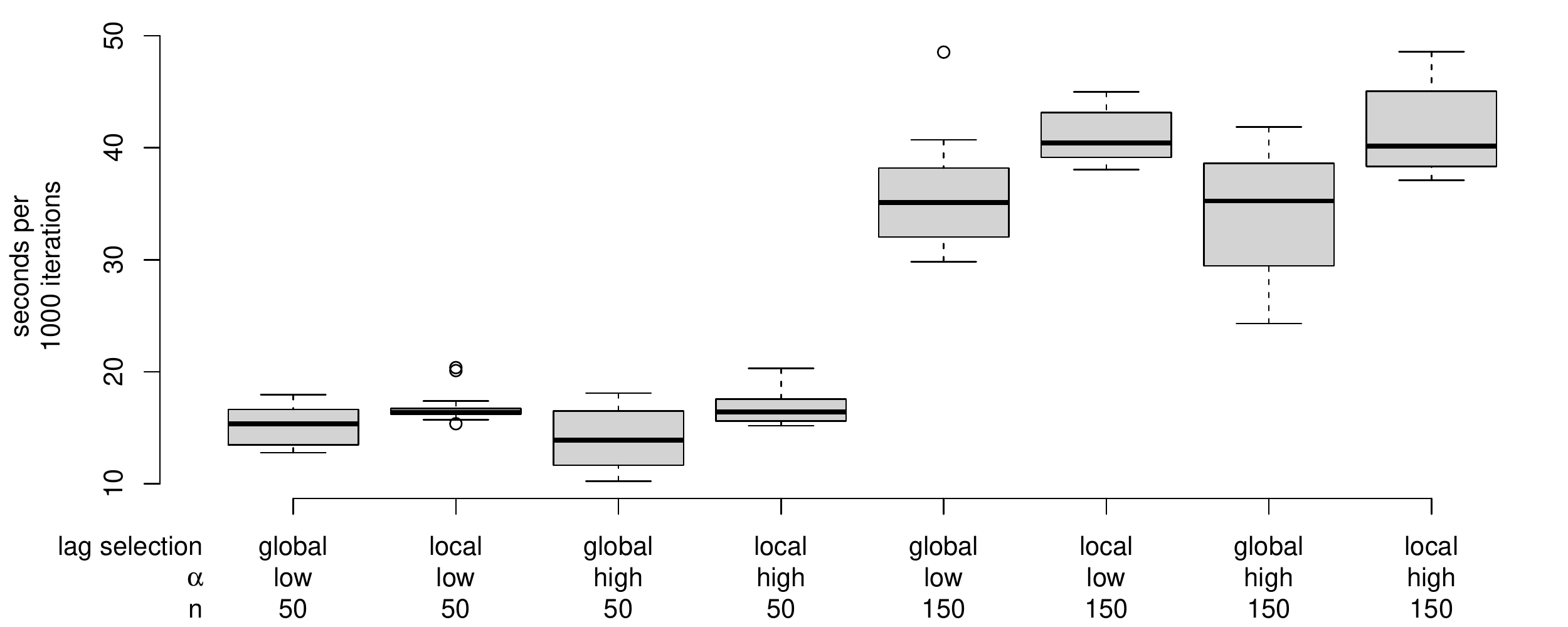}
    \end{center}
    \caption{Box plots of model run times, in seconds per 1,000 iterations of MCMC, after burn-in for all sensitivity runs on the Ricker simulation data. Groups are separated by method of lag selection, DP concentration parameter $\alpha$, and fit sample size $n$. Each group includes 16 runs.
    \label{fig:timing_boxplots}}
  \end{figure}

Figure \ref{fig:timing_boxplots} summarizes timings for all 128 runs, all of which lie between 10 and 50 seconds per 1,000 iterations. Other runs with more typical truncation levels (i.e., $H = 50$) ranged from approximately 6 to 30 seconds per 1,000 iterations. Sample size and method of lag selection are the primary influencers of running time. Constant prior lag-inclusion probabilities have appeared also to interact with low $\alpha$ to increase running time in other runs. The increasing effects of sample size and local selection are intuitive, although the burden of local selection is perhaps lower than expected. 

\subsection*{Convergence}
Among the 128 chains, 18 showed visible signs of moderate-to-high difficulty in convergence among log-likelihood or lag-selection indicators, and an additional 16 were unacceptable. Traces for $\gamma$ that remained at the truth for all inference samples were considered acceptable, while chains that switched once or very infrequently were flagged, depending on severity. None of the factors were predictive of difficulty generally, though several appeared to influence major difficulty. Chains had a greater tendency to become stuck in a mode with longer time series. Models employing global selection were more than twice as likely to experience major difficulty than those with local selection. Initialization is also important; several chains initialized with all lags off failed to select any lag during the run (i.e., longer burn-in was necessary). Mixing issues were also more common among runs with higher $\alpha$.

The remaining analyses exclude results from the 16 unacceptable chains.

\subsection*{Lag selection}
With simulated data from the Ricker model, the second lag is consistently and decisively selected. However, variations along the fourth lag can give the appearance of second-order dynamics, or aliasing can lead a model to select lag 4 only. Using posterior inclusion probability on lag 4 (for global selection, and inclusion probabilities on occupied clusters for local selection) as the response, there appears to be sensitivity to the prior inclusion probabilities. In this sense, the geometric sequence succeeds in avoiding the aliasing problem by discouraging inclusion of lag 4. Local lag selection also appears to include lag 4 less often. Fits to longer time series also included lag 4 less often in these runs. Initialization of $\bm{\gamma}$ did not sytematically affect these runs, but has appeared to contribute in other runs.  
While these appear to be common patterns, it is difficult to predict inclusion of lag 4 from run to run.

\subsection*{Number of occupied components}
Model inputs designed to influence the number of occupied mixture components are $\alpha$ and the prior signal-to-noise ratio, $\mathcal{R}$, which influences flexibility of the transition mean functional through kernel variance parameters. Higher values of each indeed increase the number of occupied components, although the influence of $\alpha$ is weak. Larger sample sizes likewise increase the number of occupied components, with a compound effect when $\mathcal{R}$ is also high.

\subsection*{Width of interval estimates}

In addition to the substantial effect of sample size in reducing the width of interval estimates, we found that low values of $\alpha$ have a similar effect. 
One possible explanation for the $\alpha$ effect is that a larger number of mixture components carry more weight a priori (through $\omega$ parameters) when $\alpha$ is high, increasing uncertainty.

\subsection*{Mean-squared error}
\begin{figure}[t!]
    \begin{center}
        \includegraphics[width=6.5in]{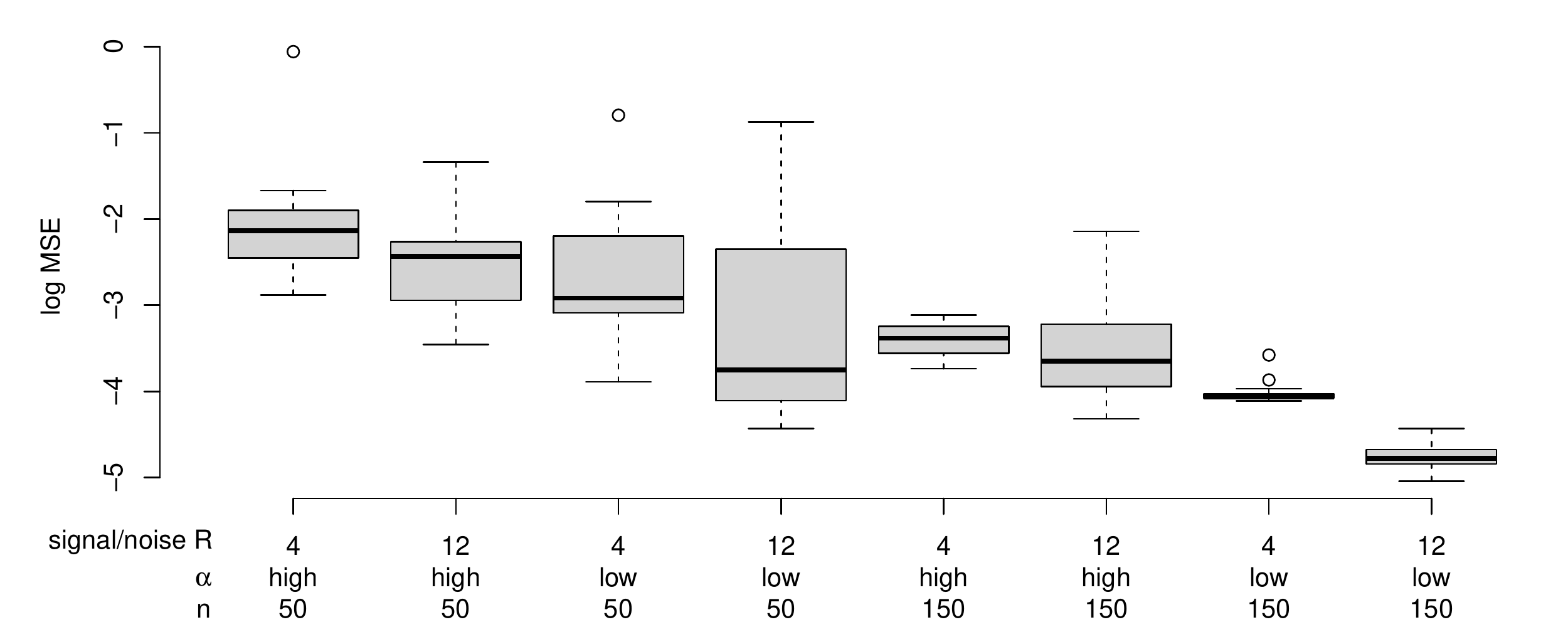}
    \end{center}
    \caption{Box plots of log mean-squared error (MSE) using sensitivity runs on the Ricker simulation data, restricted to observations where $y_{t-2} < 2.5$. Groups are separated by prior signal-to-noise $\mathcal{R}$, DP concentration parameter $\alpha$, and fit sample size $n$. Each group includes between 12 and 16 runs.
    \label{fig:mse_boxplots}}
  \end{figure}

Because the simulated transition distribution in this scenario is a nonlinear function with additive Gaussian noise, we consider estimation performance of the transition mean functional. Squared errors between the true transition function and estimated transition mean functional at each observed value were averaged across observations and a subset of posterior samples. Results are summarized in Figure \ref{fig:mse_boxplots}.

Beyond the obvious effect of sample size, forcing low $\alpha$ also decreased the mean-squared error (MSE) in these runs, likely for the same reasons it decreased the width of interval estimates. 
If we restrict attention to the estimation performance on low values of $y_{t-2}$, a region with stronger nonlinearity in the transition mean (see Figure \ref{fig:appdata_lag2only_transMean}), then the prior signal-to-noise ratio $\mathcal{R}$ becomes important, especially with the longer time series ($n=150$). Increasing $\mathcal{R}$ allows the model to better fit curvature in the transition mean function.

\section{MCMC details  for the base model} 
\label{sec:appendix_MCMC}

If we condition on the first $L$ observations of the time series, the hierarchical model (\ref{eq:hierarchicalmodel}) yields the full joint posterior distribution over all model parameters up to proportionality,
  \begin{align}
    \label{eq:fulljointpost}
    \begin{split}
    p(\cdots &\mid \{ y_t \}_{t=1}^T ) \propto \\ 
    & \prod_{t=L+1}^T \left[  \frac{ \omega_{s_t} \Ndist_{(s_t)}( \bm{y}_{t-1:L} \mid \bm{\mu}^x, \bm{\Sigma}^x) }{ \sum_{j=1}^H \omega_j \Ndist_{(j)}( \bm{y}_{t-1:L} \mid \bm{\mu}^x, \bm{\Sigma}^x) } \Ndist_{(s_t)} \left( y_t \mid  \mu^y - \sum_{\ell=1}^L \beta_{\ell}^{y} ( y_{t-\ell} - \mu_{\ell}^x ), \, \sigma^2 \right) \right] \times  \\
    & \prod_{h=1}^H \Bigg{[} \Betadist(v_h\mid 1 , \alpha)^{1_{(h<H)}}   \ p( \bm{\eta}_h \mid G_0 ) \Bigg{]} \ \Gammadist(\alpha\mid a_\alpha, b_\alpha) \ \times  \\
     & \Ndist(\bm{\beta}_0^*) \ \IWishdist\left((\bm{\Lambda}_0^*)^{-1}\right) \ \Gammadist(s_0) \ \Ndist(\bm{\mu}_0^x) \ \IWishdist((\bm{\Lambda}^{\mu_x})^{-1}) \ \times  \\
     & \prod_{r=1}^{L-1} \left[ \Ndist(\bm{\beta}_{0,r}^{\beta_x}) \ \IWishdist((\bm{\Lambda}_{0,r}^{\beta_x})^{-1}) \right] \ \prod_{\ell=1}^L \Gammadist( s_{0,\ell}^x) \, , %
     \end{split}
  \end{align}
  where
  \begin{align}
    \label{dq:eta_h_jointpost}
    \begin{split}
    p( \bm{\eta}_h \mid G_0 ) = &\Ndist\left( (\mu^y_{(h)},\bm{\beta}^y_{(h)})\mid \bm{\beta}_0^*, \sigma_{(h)}^2 (\bm{\Lambda}_0^*)^{-1} \right) \ \IGdist\left(\sigma^2_{(h)}\mid \frac{\nu_{\sigma^2}}{2}, \frac{\nu_{\sigma^2} s_0}{2} \right) \times  \\  
    & \Ndist\left(\bm{\mu}_{(h)}^x \mid \bm{\mu}_{0}^x, (\bm{\Lambda}^{\mu_x})^{-1}\right) 
    \ \prod_{r=1}^{L-1} \Ndist\left( \bm{\beta}^{x}_{r,(h)} \mid \bm{\beta}_{0,r}^{\beta_x}, (\bm{\Lambda}_{0,r}^{\beta_x})^{-1} \right) \times   \\ 
    & \prod_{\ell=1}^L \IGdist\left(\delta_{\ell,(h)}^x\mid \frac{\nu_\ell^{\delta^x}}{2}, \frac{\nu_\ell^{\delta^x} s_{0,\ell}^x}{2} \right) 
    \end{split}
  \end{align}


  \noindent The Gibbs sampler proceeds by successively sampling the parameters in the sets and manner described below.

  \subsubsection*{Latent states}

  The latent states identifying component membership for each observation $y_t$ are updated individually, for $t=L+1,\ldots,T$, using their discrete full conditional distributions $\Pr(s_t = h \mid \cdots) \propto \omega_h \Ndist_{(h)}( \bm{y}_{t-1:L} \mid \bm{\mu}^x, \bm{\Sigma}^x)  \Ndist_{(h)} \left( y_t \mid \mu^y - \sum_{\ell=1}^L \beta_{\ell}^{y} ( y_{t-\ell} - \mu_{\ell}^x ), \sigma^2 \right)$, for $h=1, \ldots, H$. The step is Metropolized by first drawing a candidate with probability mass proportional to the full conditional, excluding the current state. The Metropolis acceptance ratio is then the sum over all full conditional probabilities excluding the current state, divided by the sum over all full conditional probabilities excluding the candidate state \citep[p.\ 394]{robert2004book}.
  
  \subsubsection*{Stick-breaking weights}

  The weights $\{ \omega_h \}_{h=1}^H$ that appear in the likelihood are defined through the latent $\{ v_h \}_{h=1}^{H-1} $ which, conditional on the latent states $\{ s_t \}$ and absent the denominator in the first product term of (\ref{eq:fulljointpost}), admit $H-1$ independent beta full conditional distributions \citep{ishwarangibbs2001}. In our model, the full conditional distributions are given as 
  \begin{align}
    \label{eq:fullcond_v}
    \begin{split}
    p(\{ v_h \} \mid \cdots) &\propto \prod_{t=L+1}^T \left[ \frac{ \omega_{s_t} }{ \sum_{j=1}^H \omega_j \Ndist_{(j)}( \bm{y}_{t-1:L} \mid \bm{\mu}^x, \bm{\Sigma}^x) }  \right] \prod_{h=1}^{H-1} \Betadist(v_h\mid 1, \alpha)  \\
    &\propto \frac{ \prod_{h=1}^{H-1} \Betadist(v_h\mid 1 + n^*_h, \alpha + \sum_{k=h+1}^{H} n_k^*) }{ \prod_{t=L+1}^T \sum_{j=1}^H v_j \prod_{i=1}^{j-1} (1 - v_i) \ \Ndist_{(j)}( \bm{y}_{t-1:L} \mid \bm{\mu}^x, \bm{\Sigma}^x) } \, ,
    \end{split}
  \end{align}
  where the $n_h^* = \sum_{t=L+1}^T 1_{(s_t = h)}$, for $h=1,\ldots,H$, count membership in each of the $H$ components. We define $v_H = 1$ for convenience in notation. To obtain direct samples from this distribution, we employ the multivariate hyperrectangle slice sampler of \citet{neal2003slice} (summarized in Figure 8 of that article) to update all $v_h$, $h = 1, \ldots, H-1$, simultaneously, as follows.
  
  Let $\bm{v} = (v_1, \ldots, v_{H-1})$ denote the vector of latent beta variables used to construct $\{ \omega_h \}_{h=1}^H$, and let $g(\bm{v})$ denote the unnormalized posterior full conditional density (\ref{eq:fullcond_v}) evaluated at $\bm{v}$. 
%
%
  The algorithm employs user-specified tuning parameters $\{ \tau_h \}_{h=1}^{H-1}$, all of which we conservatively fix equal to 1.0 to ensure that the entire support of $\bm{v}$ (i.e., the hypercube $(0,1)^{H-1}$) can be reached in any iteration of MCMC.

  Let $\bm{v}^0$ denote the value of $\bm{v}$ from the previous iteration of MCMC, and $\bm{v}^1$ denote the output of this algorithm, which proceeds as follows (Figure 8 of \citealp{neal2003slice}).
  \begin{enumerate}
  
      \item Define the slice. \\
          Draw $z \sim \Unifdist ( 0, g(\bm{v}_0) ) $.
  
      \item Initialize the hyperrectangle. \\
          $\mathcal{H} = (L_1, R_1) \times \cdots \times (L_{H-1}, R_{H-1})$, where \\
          $L_h \leftarrow v^0_h - \tau_h \, U_h $, \\ 
          $R_h \leftarrow L_h + \tau_h$, \\
          with draws $U_h \simindep \Unifdist(0,1), \ \text{for} \ h = 1, \ldots, H-1$.
  
      \item Propose candidates $\bm{v}^*$ and iteratively shrink $\mathcal{H}$ when points are rejected. \\
          Repeat the following until a candidate satisfying $ z < g(\bm{v}^*) $ is found:
          \begin{enumerate}[(i)]
              \item Draw $\tilde{U}_h \simindep \Unifdist(0,1)$, for $h = 1, \ldots, H-1$.
              \item Set candidate $v^*_h \leftarrow L_h + \tilde{U}_h \, (R_h - L_h)$, for $h = 1, \ldots, H-1$.
              \item If $ z < g(\bm{v}^*)$, set $\bm{v}^1 \leftarrow \bm{v}^*$ and exit the algorithm.
              \item If $v^*_h < v^0_h$, then set $L_h \leftarrow v^*_h$, otherwise set $R_h \leftarrow v^*_h$, for \\ $h = 1, \ldots, H-1$.
          \end{enumerate}		
  \end{enumerate}

  \subsubsection*{Component-specific parameters}
  \label{sec:appendix_eta_update}

  The posterior full conditional density for each $\bm{\eta}_h$ is given by
  \begin{align}
    \label{eq:fc_eta}
    \begin{split}
    p(\bm{\eta}_h &\mid \cdots) \propto \\ 
    & \prod_{t:s_t = h} \left[  \Ndist_{(h)}( \bm{y}_{t-1:L} \mid \bm{\mu}^x, \bm{\Sigma}^x) \Ndist_{(h)} \left( y_t \, \mid \, \mu^y - \sum_{\ell=1}^L \beta_{\ell}^{y} ( y_{t-\ell} - \mu_{\ell}^x ), \, \sigma^2 \right) \right] \times  \\
    & \prod_{t=L+1}^T \left[ \sum_{j=1}^H \omega_j \Ndist_{(j)}( \bm{y}_{t-1:L} \mid \bm{\mu}^x, \bm{\Sigma}^x) \right]^{-1} \Ndist\left( (\mu^y_{(h)},\bm{\beta}^y_{(h)})\mid \bm{\beta}_0^*, \sigma_{(h)}^2 (\bm{\Lambda}_0^*)^{-1} \right) \times  \\ 
    & \IGdist\left(\sigma^2_{(h)}\mid \frac{\nu_{\sigma^2}}{2}, \frac{\nu_{\sigma^2} s_0}{2} \right) \Ndist\left(\bm{\mu}_{(h)}^x \mid \bm{\mu}_{0}^x, (\bm{\Lambda}^{\mu_x})^{-1}\right) \times \\ 
    & \prod_{r=1}^{L-1} \Ndist\left( \bm{\beta}^{x}_{r,(h)} \mid \bm{\beta}_{0,r}^{\beta_x}, (\bm{\Lambda}_{0,r}^{\beta_x})^{-1} \right) \ \prod_{\ell=1}^L \IGdist\left(\delta_{\ell,(h)}^x\mid \frac{\nu_\ell^{\delta^x}}{2}, \frac{\nu_\ell^{\delta^x} s_{0,\ell}^x}{2} \right) \, , 
    \end{split}
  \end{align} 
  for $h=1,\ldots,H$. To improve mixing of the $y$-indexed, component-specific parameters, we partition $\bm{\eta}$ into its $y$ and $x$ components $\bm{\eta}^y \equiv \{ \mu^y, \bm{\beta}^y, \sigma^2 \}$ and $\bm{\eta}^x \equiv \{ \bm{\mu}^x, \bm{\beta}_1^x, \ldots, \beta_{L-1}^x, \bm{\delta}^x \}$, and sample $p(\bm{\eta}_h \mid \cdots) = p(\bm{\eta}^x_h \mid \cdots, -\bm{\eta}^y_h) \ p( \bm{\eta}^y_h \mid \bm{\eta}^x_h, \cdots)$, where $p(\bm{\eta}^x_h \mid \cdots, -\bm{\eta}^y_h) = \int p(\bm{\eta}_h \mid \cdots) \, \diff \bm{\eta}^y_h$. This sequential sampling scheme adds little to algorithmic complexity, as the full conditional density $p(\bm{\eta}^x_h \mid \cdots)$ already contains the mixture-weight denominator $\prod_t \sum_j \omega_j \Ndist_{(j)}(\bm{y}_{t-1:L})$, precluding simple conjugate updates.

  Integrating $\bm{\eta}_h^y$ from the full conditional for $\bm{\eta}_h$ yields
  \begin{align}
    \label{eq:cc_eta_x}
    \begin{split}
    p(\bm{\eta}^x_h &\mid \cdots, -\bm{\eta}^y_h) \propto \prod_{t:s_t = h}  \Ndist_{(h)}( \bm{y}_{t-1:L} \mid \bm{\mu}^x, \bm{\Sigma}^x) \prod_{t=L+1}^T \left[ \sum_{j=1}^H \omega_j \Ndist_{(j)}( \bm{y}_{t-1:L} \mid \bm{\mu}^x, \bm{\Sigma}^x) \right]^{-1} \times  \\ 
    & \quad \Ndist\left(\bm{\mu}_{(h)}^x \mid \bm{\mu}_{0}^x, (\bm{\Lambda}^{\mu_x})^{-1}\right) \prod_{r=1}^{L-1} \Ndist\left( \bm{\beta}^{x}_{r,(h)} \mid \bm{\beta}_{0,r}^{\beta_x}, (\bm{\Lambda}_{0,r}^{\beta_x})^{-1} \right) \times  \\
    & \quad \prod_{\ell=1}^L \IGdist\left(\delta_{\ell,(h)}^x\mid \frac{\nu_\ell^{\delta^x}}{2}, \frac{\nu_\ell^{\delta^x} s_{0,\ell}^x}{2} \right) \ \det(\bm{\Lambda}^*_{1,h})^{-1/2} \, (b^*_{1,h})^{-a^*_{1,h}} \, ,
    \end{split}
  \end{align}  
  where $\bm{\Lambda}^*_{1,h} = \bm{D}_h^{'} \, \bm{D}_h + \bm{\Lambda}^*_0$; $b^*_{1,h} = \left[ \nu_{\sigma^2} s_0 + \bm{y}_{(h)}' \, \bm{y}_{(h)} + (\bm{\beta}^*_0)' \, \bm{\Lambda}^*_0 \, \bm{\beta}^*_0 - (\bm{\beta}^*_{1,h})' \, \bm{\Lambda}^*_{1,h} \, \bm{\beta}^*_{1,h} \right] / 2 $;  $a^*_{1,h} = (\nu_{\sigma^2} + n_h^*)/2 $; $\bm{\beta}^*_{1,h} = (\bm{\Lambda}^*_{1,h})^{-1}(\bm{\Lambda}^*_0 \, \bm{\beta}_0^* + \bm{D}_h^{'} \, \bm{y}_{(h)})$; $\bm{y}_{(h)}$ is a $n^*_h$-length vector containing all $y_t$ such that $s_t=h$; and $\bm{D}_h$ is a $n^*_h \times (L+1)$ design matrix whose rows correspond to $\bm{y}_{(h)}$ and are composed of $(1, \mu^x_{1,(h)} - y_{t-1}, \ldots, \mu^x_{L,(h)} - y_{t-L} )$ for each $t$ such that $s_t=h$. Note that proportionality in (\ref{eq:cc_eta_x}) is preserved with respect to the $ \{\mu^x_{\ell,(h)}\} $, which appear in the regression means for $y_t$. Aside from the factor containing normalizing weights in the mixture denominator, the full conditional for $\bm{\eta}^x_h$ could be factored into a series of conjugate updates that could serve as proposal distributions for a Metropolis step. This approach yields low acceptance rates in practice, and we instead utilize a random-walk Metropolis sampler with jointly Gaussian proposals for all parameters in $\bm{\eta}^x_h$ (with $\{ \delta^x \}$ parameters proposed on the logarithmic scale), which are evaluated using (\ref{eq:cc_eta_x}). Proposals that produce computationally singular covariance matrices are automatically rejected.

  The full conditional distribution for $\bm{\eta}^y_h$ factors as $p(\sigma^2_h \mid \cdots, -\bm{\beta}_h^* ) \ p(\bm{\beta}_h^* \mid \sigma^2_h, \cdots)$ and is drawn sequentially as $\sigma^2_h \mid \cdots, -\bm{\beta}_h^* \sim \IGdist(a_{1,h}^*, b_{1,h}^*)$ and $\bm{\beta}_h^* \mid \sigma^2_h, \cdots \sim \Ndist\left( \bm{\beta}^*_{1,h}, \sigma^2_h \, (\bm{\Lambda}^*_{1,h})^{-1} \right)$.

  \subsubsection*{Parameters in the base measure}

  Let $n^* = \sum_{h=1}^H 1_{(n_h^* > 0)}$ count the total number of occupied components. The posterior conditional density for $\bm{\beta}_0^*$ is proportional to \\ $\prod_{\{h:n_h^* > 0\}}\left[ \Ndist(\bm{\beta}_h^* \mid \bm{\beta}_0^*, \sigma_h^2 \, (\bm{\Lambda}_0^*)^{-1} ) \right] \Ndist(\bm{\beta}_0^*\mid \bm{b}_0^*, \bm{S}_0^*) $, yielding a Gaussian update with covariance matrix $\bm{S}_1^* = \left( \sum_{\{h:n_h^*>0\}} \sigma_h^{-2} \, \bm{\Lambda}_0^* + (\bm{S}_0^*)^{-1} \right)^{-1}$ and mean \\ $\bm{S}_1^* \left( (\bm{S}_0^*)^{-1} \, \bm{b}_0^* + \bm{\Lambda}_0^* \sum_{\{h:n_h^* > 0\}} \sigma_h^{-2} \, \bm{\beta}_h^*  \right) $.

  The posterior conditional density for $(\bm{\Lambda}_0^*)^{-1}$ is proportional to \\ $\prod_{\{h:n_h^* > 0\}}\left[ \Ndist(\bm{\beta}_h^* \mid \bm{\beta}_0^*, \, \sigma_h^2 \, (\bm{\Lambda}_0^*)^{-1} ) \right] \, \IWishdist((\bm{\Lambda}_0^*)^{-1}\mid \nu^*, \, \nu^* \, \bm{\Psi}_0^*)$, yielding an inverse-Wishart update with degrees of freedom $\nu^* + n^*$ and scale matrix \\ $\nu^* \, \bm{\Psi}_0^* + \sum_{\{h:n_h^* > 0\}} \sigma_h^{-2} \, (\bm{\beta}_h^* - \bm{\beta}_0^*) \, (\bm{\beta}_h^* - \bm{\beta}_0^*)'$.

  The posterior conditional density for $s_0$ is proportional to \\ $\prod_{\{h : n_h^* > 0 \}} \left[ \IGdist( \nu_{\sigma^2}/2, \, \nu_{\sigma^2} \, s_0/2) \right] \Gammadist(s_0\mid a_{s_0}, \, b_{s_0})$, yielding a gamma update with shape $a_{s_0} + \nu_{\sigma^2} \, n^*/2$ and rate $b_{s_0} + \nu_{\sigma^2} \sum_{\{h:n_h^* > 0\}} \sigma_h^{-2} / 2 $. Updates for $\{ s_{0,\ell}^x \}$ are analogous, with $\delta^x_{\ell, (h)}$ replacing $\sigma_h^2$, except that all $H$ values are required for each update.

  All remaining parameters in the base measure have standard conditionally conjugate updates. Because all $\{ \bm{\eta}_h^x \}$ parameters are used in the local $q_h(\bm{x})$ weights, the updates for associated $G_0$ parameters require all $H$ values, rather than the $n^*$ values associated with occupied components. 

  \subsubsection*{DP concentration parameter}

  The posterior full conditional density for the DP concentration parameter $\alpha$ is proportional to $\prod_{h=1}^{H-1} \left[ \Betadist(v_h\mid 1, \alpha) \right] \Gammadist(\alpha\mid a_\alpha, b_\alpha) $, yielding a gamma update with shape $a_\alpha + H - 1$ and rate $b_\alpha - \log(\omega_H)$.

  \subsubsection*{Adaptation}

After initialization, MCMC begins with a tuning phase for diagonal elements of the covariance matrix in the candidate-generating Gaussian proposal distribution for $\{ \bm{\eta}^x_h \}_{h=1}^H $. Optionally, an adaptation phase may then be used to further tune a full candidate covariance matrix. This proceeds in four steps. In the first step, the initial covariance matrix is globally scaled to adjust acceptance rates collected over a short run. This is repeated iteratively until all acceptance rates fall within a pre-specified range (we set the range low, e.g., $[0.02, 0.20]$, to promote exploration) or a maximum number of attempts is reached. In the second step, the proposal variances are scaled locally by parameter groups corresponding to $\bm{\mu}^x$, $\{ \bm{\beta}^x_r \}$, and $\bm{\delta}^x$, while preserving correlations. In the third step, empirical cross-covariance matrices are estimated from a longer run. In the final step, these empirical covariance matrices are scaled globally until acceptance rates fall within the pre-specified range, or a maximum number of attempts is reached. At this point, adaptation ceases and the scaled empirical covariance matrices are used for subsequent random-walk proposals. We advise against adapting prematurely, which can cause the chain to develop an affinity for a local mode during burn-in.

\section{Details for lag selection} 
\label{sec:appendix_lagSelection}

\subsection{Posterior inference with global selection}
\label{sec:appendix_lagSelection_globalPost}

Conditional on $\bm{\gamma} = (\gamma_1, \ldots, \gamma_L)$, the selection effect on the mixture kernels can be passed through to the $\bm{D}_h$ matrices, for which all elements in column $\ell+1$ are replaced with 0s if $\gamma_\ell = 0$. Because $\bm{\eta}^x_h$ is updated with a Metropolis step, one simply draws candidate values and evaluates (\ref{eq:cc_eta_x}) with each $\Ndist_{(h)}(\bm{y}_{t-1:L})$ for $h = 1, \ldots, H$, and $t=L+1, \ldots, T$, appropriately modified (with respect to $\bm{\gamma}$). The full conditional distribution for $\bm{\eta}^y_h$ is then sampled using the modified $\bm{D}_h$. All other updates proceed as before, using the appropriately modified $\Ndist_{(h)}(\bm{y}_{t-1:L})$ and kernel means.

The posterior full conditional probability that $\gamma_\ell = 1$ is
\begin{align}
  \label{eq:fc_gamma}
  \Pr(\gamma_\ell = 1 \mid \cdots ) = \frac{ \pi^{\gamma}_\ell a^\gamma_\ell }{ \pi^{\gamma}_\ell a^\gamma_\ell + (1-\pi^{\gamma}_\ell)  b^\gamma_\ell } \, ,
\end{align}
where \\ $a^\gamma_\ell = \prod_{t=L+1}^T \left[ \Ndist_{(s_t)}( \bm{y}_{t-1:L} \mid \gamma_\ell = 1 ) \, \Ndist_{(s_t)}( y_t \mid \gamma_\ell = 1 ) \left( \sum_{j=1}^H \omega_j  \Ndist_{(j)}( \bm{y}_{t-1:L} \mid \gamma_\ell = 1 )  \right)^{-1} \right] $ \\ and \\
$b^\gamma_\ell = \prod_{t=L+1}^T \left[ \Ndist_{(s_t)}( \bm{y}_{t-1:L} \mid \gamma_\ell = 0 ) \, \Ndist_{(s_t)}( y_t \mid \gamma_\ell = 0 ) \left( \sum_{j=1}^H \omega_j  \Ndist_{(j)}( \bm{y}_{t-1:L} \mid \gamma_\ell = 0 )  \right)^{-1} \right] $. \\ The Gaussian kernel densities in these expressions are modified to reflect either $\gamma_\ell = 1$ or 0, and appropriately reflect all other $\{ \gamma_k \}_{k \ne \ell}$.

Instead of drawing from the individual full conditionals, we update $\bm{\gamma}$ as a block using a collapsed conditional, with $ \{ \bm{\eta}^y_h \} $ integrated out, as in Supplement \ref{sec:appendix_eta_update}. First, a number of proposed switches, $k \in \{1, 2, 3\}$, is drawn from a truncated geometric distribution. Then, a uniformly drawn subset of $k$ indices among $\{1, \ldots, L\}$, denoted $\mathcal{K}$, identifies which elements of the proposed $\bm{\gamma}^{\mbox{cand}}$ are switched from the current state (individually, from 0 to 1 or from 1 to 0). Symmetry of this proposal distribution (\citealp[Sec.\ 3.3]{schafer2013sequential}) yields the Metropolis ratio, $p( \bm{\gamma}^{\mbox{cand}} \mid \cdots, - \{ \bm{\eta}^y_h \} ) / p( \bm{\gamma}^{\mbox{old}} \mid \cdots, -\{ \bm{\eta}^y_h \} )$, where
\begin{align}
  \label{eq:gamma_global_cc}
  p( \bm{\gamma}_\mathcal{K} \mid \cdots, -\{ \bm{\eta}^y_h \} ) \propto &\prod_{k \in \mathcal{K}} ( \pi_k^\gamma )^{\gamma_k} ( 1 - \pi_k^\gamma )^{1 - \gamma_k} \times \prod_{t=L+1}^T \left[ \Ndist_{(s_t)}( \bm{y}_{t-1:L} ) \left( \sum_{j=1}^H \omega_j \, \Ndist_{(j)}( \bm{y}_{t-1:L} ) \right)^{-1} \right] \times \nonumber \\ &\prod_{h=1}^H \det( \bm{\Lambda}^*_{1,h} )^{-1/2} 
  \, (b^*_{1,h})^{-a^*_{1,h}} \, ,
\end{align}
with all quantities calculated using the full $\bm{\gamma}$ vector under evaluation.

\subsection{Transition density estimation under global lag selection}
\label{sec:appendix_lagSelection_globalFunctionals}
For any value of $y$ and $\bm{x}$, or over a multidimensional grid of values, samples for $f_{Y|X}$ are calculated from 
\begin{align}
  \label{eq:condmodel_trunc_lagselect}
  \tilde{f}_{Y|X}(y\mid\bm{x}, \bm{\gamma}) =  \sum_{h=1}^H \tilde{q}_h(\bm{x} \mid \bm{\gamma}) \, \Ndist_{(h)}(y\mid \mu(\bm{x} \mid \bm{\gamma}), \sigma^2 ) \, ,
\end{align}
with $\tilde{q}_h(\bm{x} \mid \bm{\gamma}) = \omega_h \Ndist_{(h)}(\bm{x} \mid \bm{\gamma}) / \sum_{j=1}^H \omega_j \Ndist_{(j)}(\bm{x} \mid \bm{\gamma})$ and $\mu(\bm{x} \mid \bm{\gamma}) = \mu^y - \sum_{\ell=1}^L \gamma_\ell \, \beta^y_\ell \, (x_\ell - \mu_\ell^x)$. The samples can then be used to create pointwise estimates and intervals for $\tilde{f}_{Y|X}$. The expression for the transition mean becomes $\tilde{\E}_{Y|X}(y\mid\bm{x}, \bm{\gamma}) = \sum_{h=1}^H \tilde{q}_h(\bm{x} \mid \bm{\gamma}) \, \mu_{(h)}(\bm{x} \mid \bm{\gamma})$. Analogous expressions include $\bm{\gamma}$ in the procedure for estimating quantiles in Section \ref{sec:transdens_estimation}. Likewise, $K$-step-ahead forecasts are inductively sampled with $(s, y)_{T+k}$ pairs for $k = 1, \ldots, K$, following the first two levels of the hierarchical model (\ref{eq:hierarchicalmodel}), adjusted for $\bm{\gamma}$, for each posterior sample. While dependence on other parameters in (\ref{eq:hierarchicalmodel}) is implicit in the preceding expressions, we add explicit dependence on $\bm{\gamma}$ in order to emphasize the modifications necessary to include lag dependence.

\subsection{Posterior inference with local selection}
\label{sec:appendix_lagSelection_localPost}

The posterior full conditional probability that $\gamma^{(h)}_\ell = 1$ is
\begin{align}
  \label{eq:fc_gamma_h}
  \Pr \left( \gamma^{(h)}_\ell = 1 \mid \cdots \right) = \frac{ \pi^{\gamma}_\ell a^\gamma_{h,\ell} }{ \pi^{\gamma}_\ell a^\gamma_{h,\ell} + (1-\pi^{\gamma}_\ell) b^\gamma_{h,\ell} } \, ,
\end{align}
where \\ $a^\gamma_{h,\ell} = \prod_{ \{ t:s_t = h \} } \left[ \Ndist_{(h)}( \bm{y}_{t-1:L} \mid \gamma^{(h)}_\ell = 1 ) \, \Ndist_{(h)}( y_t \mid \gamma^{(h)}_\ell = 1 ) \right] \, \prod_t  \left( \sum_{j=1}^H \omega_j  \Ndist_{(j)}( \bm{y}_{t-1:L} \mid \gamma^{(h)}_\ell = 1 )  \right)^{-1} $ \\ and \\
$b^\gamma_{h,\ell} = \prod_{ \{ t:s_t = h \} } \left[ \Ndist_{(h)}( \bm{y}_{t-1:L} \mid \gamma^{(h)}_\ell = 0 ) \, \Ndist_{(h)}( y_t \mid \gamma^{(h)}_\ell = 0 ) \right] \, \prod_t \left( \sum_{j=1}^H \omega_j  \Ndist_{(j)}( \bm{y}_{t-1:L} \mid \gamma^{(h)}_\ell = 0 )  \right)^{-1} $. \\ Note that the first product in each line is over all time points allocated to mixture component $h$, and the second product is over all time points $t = L+1, \ldots, T$. The Gaussian densities in these expressions are from (\ref{eq:fulljointpost}), modified to reflect either $\gamma^{(h)}_\ell = 1$ or $0$, and appropriately reflecting all other $\{ \gamma^{(i)}_k : i \ne h\, \text{or}\, k \ne \ell \}$, which are held constant. 


Instead of drawing from the individual full conditionals, we update each $\bm{\gamma}_h$ as a block using collapsed conditionals, with $ \{ \bm{\eta}^y_h \} $ integrated out, as in Supplement \ref{sec:appendix_eta_update}. First, a number of proposed switches, $k \in \{1, 2, 3\}$, is drawn from a truncated geometric distribution. Then, a uniformly drawn subset of $k$ indices among $\{1, \ldots, L\}$, denoted $\mathcal{K}$, identifies which elements of the proposed $\bm{\gamma}^{\mbox{cand}}_h$ are switched from the current state (individually, from 0 to 1 or from 1 to 0). Symmetry of this proposal distribution (\citealp[Sec.\ 3.3]{schafer2013sequential}) yields the Metropolis ratio, $p( \bm{\gamma}^{\mbox{cand}}_h \mid \cdots, - \{ \bm{\eta}^y_h \} ) / p( \bm{\gamma}^{\mbox{old}}_h \mid \cdots, -\{ \bm{\eta}^y_h \} )$, where
\begin{align}
  \label{eq:gamma_local_cc}
  \begin{split}
  p( \bm{\gamma}_{h, \mathcal{K}} \mid \cdots, -\{ \bm{\eta}^y_h \} ) \propto &\prod_{k \in \mathcal{K}} ( \pi_k^\gamma )^{\gamma_k^{(h)}} ( 1 - \pi_k^\gamma )^{1 - \gamma^{(h)}_k} \times  \prod_{t : s_t = h} \Ndist_{(h)}( \bm{y}_{t-1:L} ) \, \times \\ 
  & \prod_{t=L+1}^T \left[ \left( \sum_{j=1}^H \omega_j \, \Ndist_{(j)}( \bm{y}_{t-1:L} ) \right)^{-1} \right] \, \det( \bm{\Lambda}^*_{1,h} )^{-1/2} 
  \, (b^*_{1,h})^{-a^*_{1,h}} \, ,  \end{split}
\end{align}
with all quantities calculated using both the full $\bm{\gamma}_h$ vector under evaluation, and all other $\{ \bm{\gamma}_j : j \ne h \}$ held constant.

Updates for $\pi^{\gamma}_\ell$ are presented in \citet{chung2009probitsb_ssvs}, and proceed as follows. Introduce $\xi_\ell = 1_{(\pi^{\gamma}_\ell > 0)}$, for $\ell = 1, \ldots, L$. Then, conditional on $\xi_\ell = 1$, draw $\pi^{\gamma}_\ell \sim \Betadist \left( a^{\pi}_\ell + \sum_{h=1}^H \gamma_{\ell}^{(h)} , b^{\pi}_\ell + H - \sum_{h=1}^H \gamma_{\ell}^{(h)} \right)$; otherwise, set $\pi^{\gamma}_\ell=0$. Then, to update each $\xi_\ell$, set $\xi_\ell = 1$ if $\sum_{h=1}^H \gamma_{\ell}^{(h)} > 0$; otherwise, the full conditional probability that $\xi_\ell = 1$ is $\alpha^\pi_\ell / ( \alpha^\pi_\ell - 1 + 1/\pi^\pi_\ell )$, where $\alpha^\pi_\ell = \Gamma( b^\pi_\ell + H ) \Gamma( a^\pi_\ell + b^\pi_\ell ) / \left[ \Gamma( b^\pi_\ell ) \Gamma( a^\pi_\ell + b^\pi_\ell + H ) \right] $.

\subsection{Illustration of global selection on linear autoregressive simulation}
\label{sec:appendix_lagSelection_AR2}

We demonstrate the model's ability to identify simple structure, for which the proposed model is over-specified. Although each of the nonlinear, non-Gaussian, and mixture capabilities are not necessary in this case, the model performs well. A stationary time series was generated from the model 
\begin{align*}
  y_t = \mu + \phi_1(y_{t-1} - \mu) + \phi_2(y_{t-2} - \mu) + \epsilon_t \, , \qquad \epsilon_t \simiid \Ndist(0, \sigma^2) \, ,
\end{align*}
with $\mu = 2.5$, $\phi_1 = 1.2$, $\phi_2 = -0.7$, and $\sigma^2 = 1.0$. We then fit the proposed nonparametric model to series of length $T=75$ and $T=305$ with a lag horizon of $L=5$ (so that 70 and 300 observations contribute to the likelihood), DP truncation at $H=25$, and default prior settings. With the short time series, both global and local lag selection recover the true structure, with all chains decisively selecting the first two lags only. With the long time series, all methods select the first two lags and include lag 3 in a non-negligible fraction of the samples (ranging from 0.2 to 0.6, with reasonable mixing in each chain). Other inferences from all fits appear to accurately recover the truth, with exception that a few runs with local selection occasionally tend to over-fit the data, adding one or two unnecessary mixture components.

\begin{figure}[p]
  \begin{center}
    \begin{tabular} {ll}
      \includegraphics[width=2.75in, page=1]{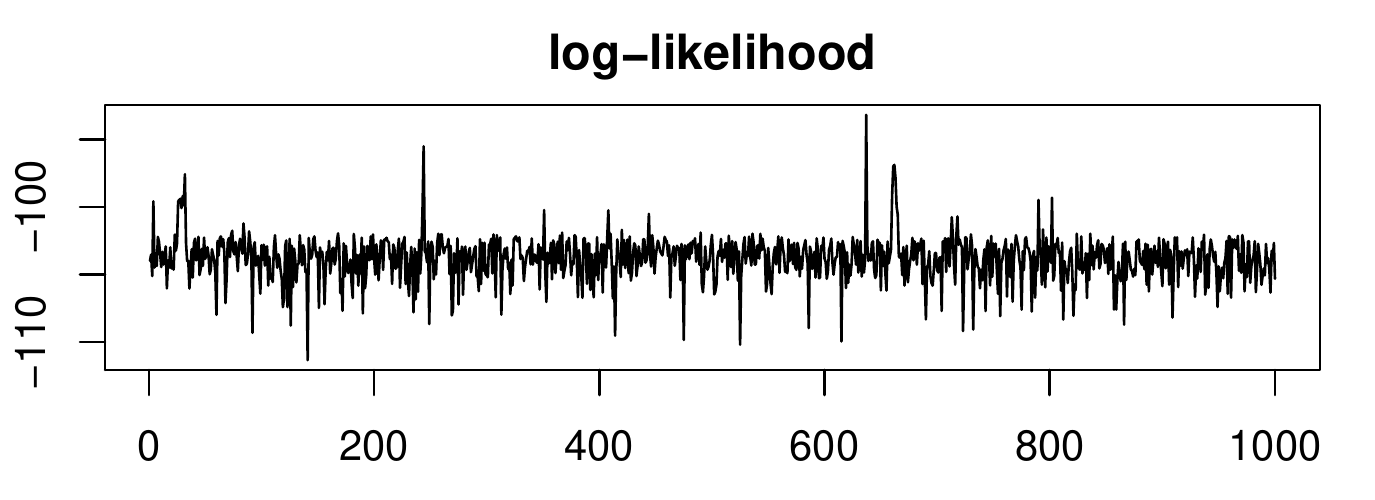} & \hspace{-0.15in}
      \includegraphics[width=2.75in, page=2]{figures/Rsimdata_AR2seed1_n70_K5_SigXdiag_VSglobal_gaminitall_snr5_1121191_paper.pdf} \\ 
      \includegraphics[width=2.75in, page=5]{figures/Rsimdata_AR2seed1_n70_K5_SigXdiag_VSglobal_gaminitall_snr5_1121191_paper.pdf} & \hspace{-0.15in}
      \includegraphics[width=2.75in, page=6]{figures/Rsimdata_AR2seed1_n70_K5_SigXdiag_VSglobal_gaminitall_snr5_1121191_paper.pdf} \\
      \includegraphics[width=2.75in, page=7]{figures/Rsimdata_AR2seed1_n70_K5_SigXdiag_VSglobal_gaminitall_snr5_1121191_paper.pdf} & \hspace{-0.15in}
      \includegraphics[width=2.75in, page=8]{figures/Rsimdata_AR2seed1_n70_K5_SigXdiag_VSglobal_gaminitall_snr5_1121191_paper.pdf} \\
      \includegraphics[width=2.75in, page=30]{figures/Rsimdata_AR2seed1_n70_K5_SigXdiag_VSglobal_gaminitall_snr5_1121191_paper.pdf} & \hspace{-0.15in}
      \includegraphics[width=2.75in, page=20]{figures/Rsimdata_AR2seed1_n70_K5_SigXdiag_VSglobal_gaminitall_snr5_1121191_paper.pdf} \\ 
      \includegraphics[width=2.75in, page=22]{figures/Rsimdata_AR2seed1_n70_K5_SigXdiag_VSglobal_gaminitall_snr5_1121191_paper.pdf} & \hspace{-0.15in}
      \includegraphics[width=2.75in, page=24]{figures/Rsimdata_AR2seed1_n70_K5_SigXdiag_VSglobal_gaminitall_snr5_1121191_paper.pdf} \\
      \includegraphics[width=2.75in, page=16]{figures/Rsimdata_AR2seed1_n70_K5_SigXdiag_VSglobal_gaminitall_snr5_1121191_paper.pdf} & \hspace{-0.15in}
      \includegraphics[width=2.75in, page=18]{figures/Rsimdata_AR2seed1_n70_K5_SigXdiag_VSglobal_gaminitall_snr5_1121191_paper.pdf} \\
    \end{tabular}
    
  \end{center}
  \caption{MCMC trace plots for the model fit to 70 observations of the simulated second-order autoregression with global lag selection. Red horizontal lines indicate true parameter values. In the lag-inclusion plots (second and third row), {\it p} refers to the Monte Carlo estimate of the posterior probability of inclusion for that lag. 
  \label{fig:ar2plots}}
\end{figure}

Figure \ref{fig:ar2plots} provides trace plots for key quantities from one model fit to 70 observations with global selection, including the log-likelihood, number of occupied components, selection indicators for the first four lags, the observation (innovation) variance for the most populated component, the first three $\beta^y$ coefficients for the most populated component, the center $\mu^y$ for the most populated component, and the intercept $(\mu^y + \sum_{\ell=1}^L \gamma_\ell \, \beta_\ell^y \, \mu_\ell^y)$ for the most populated component, thinned to 1,000 samples. The trace for the log-likelihood indicates that the chain is no longer traversing across substantially different component configurations. 
Most observations belong to one component throughout MCMC.
Lags 1 and 2 are on for all inference samples, while the remaining lags are off for nearly all inference samples.
Trace plots for the kernel parameters faithfully track the true values (indicated with horizontal red lines in the plots). Note that the sign is switched for the coefficients in the model formulation. The only trace without a precise marginal posterior distribution in this chain is for $\mu^y$, which in the model is replaced by $\mu_\ell^x$ parameters in the lag summands, and thus over-parameterized for this stationary time series. However, the intercept, which is a function of $\mu^y$ and lag coefficients is precisely identified. Trace plots for the coefficients of lags 4 and 5 are similar to that of lag 3, reflecting their prior with mean 0 in the next level of the hierarchy.

Inferences for the transition mean surface and transition densities for specific lag values (not shown), both as functions of $y_{t-1}$ and $y_{t-2}$, are consistent with the data-generating mechanism. Specifically, the estimated mean surface is very close to the true plane. Posterior mean estimates of transition densities are nearly the correct Gaussian distributions. Furthermore, marginal posterior standard deviations are nearly identical to standard errors from a correctly specified linear model fit to the time series. Hence, conditional on admittedly overconfident lag inferences, the proposed model performs well in a simple scenario, with surprisingly low cost for additional flexibility.

\end{document}